\documentclass[a4paper,fleqn,usenatbib]{mnras}

\usepackage[T1]{fontenc}
\usepackage{ae,aecompl}

\usepackage{graphicx}	
\usepackage{amsmath}	
\usepackage{amssymb}	
\usepackage{threeparttable}
\usepackage[justification=centering]{caption}

\title[Internal Kinematics of UM 461 and CTS 1020]{Internal Kinematics of UM 461 and CTS 1020}

\author[M. S. Carvalho et al.]{
Maiara S. Carvalho,$^{1}$\thanks{E-mail: mscarvalho@astro.ufsc.br (MSC); plana@uesc.br (HP)}
and Henri Plana$^{2}$ 
\\
$^{1}$Departamento de F\'isica-CFM, Universidade Federal de Santa Catarina, 88040-900, Florian\'opolis, SC, Brazil\\
$^{2}$Laborat\'orio de Astrof\'isica Te\'orica e Observacional, Universidade Estadual de Santa Cruz, 45650-000, Ilh\'eus, BA, Brazil\\
}

\date{Accepted XXX. Received YYY; in original form ZZZ}

\pubyear{2018}

\begin{document}
\label{firstpage}
\pagerange{\pageref{firstpage}--\pageref{lastpage}}
\maketitle

\begin{abstract}
We have used integral field spectroscopy to study the internal kinematics of the H\,{\sc ii} galaxies CTS 1020 and UM 461. We based our analysis on the  velocity and velocity dispersion maps, and intensity-velocity dispersion ($I-\sigma$) and velocity-velocity dispersion ($V_r-\sigma$) diagrams. We found that the motion in both star-forming knots of UM 461 has different patterns, suggesting a weak kinematical connection between the knots. The overall kinematics of the galaxy is probably affected by stellar feedback. CTS 1020 has an ordered motion with a gradient compatible with a disc rotating at $\sim50$ km s$^{-1}$, though the velocity field is disturbed. In both galaxies the highest and lowest $\sigma$ values are distributed in the outer parts and are associated with the diffuse gas that permeates the galaxies. UM 461 has a ring-like structure with small regions of increasing $\sigma$ in the eastern knot, which resemble what we could expect in a collect and collapse scenario of star formation. We found that UM 461 seems to be more susceptible to stellar feedback, whereas in CTS 1020 the gravitational potential dominates.
\end{abstract}




\section{Introduction}
H{\,\sc ii} galaxies, also known as \it blue compact dwarf \rm depending on classification criteria \citep{Melnick1985}, are a subclass of dwarf galaxies characterized by its compactness ($\sim1$ kpc), high star formation rate and a spectrum dominated by intense emission lines superimposed to a weak stellar continuum, which resembles that observed in giant H{\,\sc ii} regions in spiral galaxies \citep{Sargent1970}. Among the observed emission lines are the hydrogen Balmer series and the forbidden lines of oxygen ([O\,{\sc iii}] $\lambda\lambda4959$, 5007, 4363, [O\,{\sc ii}] $\lambda\lambda3726$, 3729), nitrogen ([N\,{\sc ii}] $\lambda\lambda6548$, 6583), and sulfur ([S\,{\sc ii}] $\lambda\lambda6716$, 6731). 

The fact of being gas-rich and metal-poor objects raised the hypothesis that these galaxies were young systems forming their first stars \citep{Searle1972},
but the idea of being old systems with intermittent star formation bursts interleaved by quiescent periods has been supported by observations of an underlying old stellar population \citep{Thuan1983,Telles1997a,Westera2004,Corbin2006}. 

An important characteristic of these galaxies is the supersonic nature of their emission line profile, which is broader than that observed in typical H{\,\sc ii} regions \citep{Smith1970,Smith1971}.
\citet{Terlevich1981} proposed that the gravitational potential is responsible for this supersonic line widths, as they found a correlation between H$\beta$ luminosity and velocity dispersion ($L\sim\sigma^{4}$) and radius and velocity dispersion ($R\sim\sigma^{2}$) in giant H{\,\sc ii} regions. This scenario was further supported by \citet{TenorioTagle1993}, who proposed a model to explain and maintain the supersonic motion, which is given by the constant passage of low-mass stars producing bow shocks. 
Alternatively, the supersonic motions may be maintained by the mechanical energy injected into the interstellar medium from the ongoing star formation activity, stellar winds, radiation pressure and supernovae explosions, all of which contribute to increase the turbulence \citep{Green2010,Moiseev2012,Moiseev2015}. These scenarios have difficulties to explain the $L\sim\sigma^{4}$ and $R\sim\sigma^{2}$ relations.

Which of these mechanisms is dominant remains an open problem. \citet{Gallagher1983} suggested that the dominant mechanism depends on the system scale, being the gravitational potential in regions of hundreds of parsecs (supergiant H{\,\sc ii} regions) and energy from massive young stellar populations in regions of tens of parsecs (giant H{\,\sc ii} regions).
Furthermore, \citet{TenorioTagle1996} demonstrated that the mechanical energy from massive stars is not sufficient to explain the observed line broadening, in agreement with \citet{Yang1996} who found that the stellar winds and supernova explosions act increasing the dispersion caused by the gravitational potential.

Kinematics and dynamics of H{\,\sc ii} galaxies have been investigated, early on, by looking at their velocity dispersion. Rapidly a scale relation between emission line and velocity dispersion have been established \citep{Melnick1988}. These scale relations have been used early on as distance indicator. Taking advantage of larger survey and better data, recent studies could minimise errors in these relations \citep{Bordalo2011, Chavez2014}. These studies also show that such relations are subject to evolutionary effects, responsible, according to authors, for part of the dispersion of such relations. More recently, high redshift H{\,\sc ii} galaxies have been used as tools for precise cosmology \citep{Terlevich2015}.

Kinematical studies of H\,{\sc ii} galaxies with 2D mapping instrumentation were first made by \citet{Ostlin1999,Ostlin2001} and
focused on the analysis of velocity field to determine the mass distribution using rotation curves.
The results showed a disturbed velocity field and supersonic velocity dispersion in a small sample of H\,\sc ii \rm galaxies, what pointed out that velocity dispersion dominates the gravitational potential. However, as the morphology of the galaxies suggested interaction or merger, the authors concluded that those galaxies were not systems in equilibrium, but in a merger process \citep{Ostlin2001}. Recently, using integral field spectroscopy, \citet{Lagos2016} and \citet{Kumari2017} came up with the same conclusion for the galaxies Tol 65 and NGC 4449, respectively, with a merger being responsible for trigger the star formation. The kinematics of both objects seems to be affected by stellar feedback \citep{Lagos2016,Kumari2017}.
Further evidence of stellar feedback on H\,\sc ii \rm galaxies is presented by \citet{Cairos2017a,Cairos2017b} that found supersonic velocity dispersion in areas surrounding H\,\sc ii \rm regions and in the outskirts of the galaxies.

In order to disentangle the line broadening mechanisms, we
used diagnostic diagrams, such as $I - \sigma $, $I - V_r$ and $V_r - \sigma$, that have been revealed to be precious tools to find signatures of peculiar motions, as expanding shells, radial motions \citep{MunozTunon1996,Bordalo2009,Plana2017} and of turbulent ISM and \hbox{H\,{\sc ii}} region \citep{Moiseev2012}.

The paper is organized as follows. In Section~\ref{sec:OR} we present observations  and  the reduction techniques used.  The ionized gas kinematics is presented  in \S~\ref{sec:IFUresults}, discussion of the diagnostic diagrams of H\,{\textsc{ii}} complexes is presented in \S~\ref{sec:diagrams} along with a statistical analysis of specific diagrams in section \S~\ref{sec:statistics}. The section \S~\ref{sec:PCA} is dedicated to a Principal Component Analysis of the data cubes.
In  \S~\ref{sec:discussion} our results are discussed. Finally in  \S~\ref{sec:conclusion}, we give the summary and draw general conclusion.

\section{Sample, Observations and Data Reduction} \label{sec:OR}

\subsection{Sample}

\subsubsection{UM 461}

UM 461 was discovered in objective prism survey from the University of Michigan \citep{MacAlpine1981}.
It has an optical structure formed by two compact star-forming knots, enveloped by a diffuse medium distorted toward the southwest, as seen in Fig.\ \ref{Fig01}. 
The knots are off-center, the brightest located east of the galaxy. 
The southwest extension of the nebulosity has been attributed to tidal effect due to an interaction with UM 462 \citep{Lagos2011,Noeske2003,Doublier1999}. 
\citet{Taylor1995} first studied this interpretation as they observed distortions in the isophotes of H\,{\sc i} emission of UM 461 towards UM 462, suggesting a binary system in interaction. 
On the other hand, in \hbox{H\,{\sc i}} observations from \citet{Zee1998} 
these two galaxies appear as isolated systems with no clear interaction signs. 
Although a possible interaction with UM 462 could be responsible for trigger the star formation in UM 461, \citet{OlmoGarcia2017} show evidence of accretion of metal-poor gas in the latter, producing a difference in metallicity along the galaxy and probably fueling the star formation activity.

Near-infrared observations showed that the western knot can be spatially resolved into several stellar clusters and complexes, whereas the eastern knot is more compact \citep{Noeske2003,Lagos2011}. These clusters are young ($\sim 5$ Myr) with diameters smaller than 37 pc and masses of the order of $\sim10^4-10^6\,M_\odot$. A stellar population with ages $\gtrsim10^8$ yr is embedded in the nebular component that envelops the star-forming knots \citep{Lagos2011}.
In a recent work, \citet{Lagos2018} report VIMOS-IFU observations of UM 461. They conclude that the ISM is well mixed, at large scale, but their study also reports a low metallicity region close to the brightest H{\,\sc ii} region. It leads the authors to imagine a scenario where a recent infall of a low mass metal-poor dwarf or H{\,\sc i}  cloud occurred.

\subsubsection{CTS 1020}

CTS 1020 was also discovered by objective prism technique in the Cal\'an-Tololo Survey \citep{Maza1991}. It has a roughly spheroidal form and a well defined nucleus (Fig.\ \ref{Fig01}), classified as a \hbox{H\,{\sc ii}} type II according to \citet{Telles1997b}.
It is a less studied galaxy, but its observational properties are reported in a number of catalogs \citep{Maza1991,Kilkenny1997,Stobie1997,Kehrig2004,Lagos2007,Kopparapu2008,Jones2009}.
Analysis of H$\beta$ equivalent width indicates an intense star formation activity in a region slightly shifted from the center of the galaxy \citep{Lagos2007}.

\subsection{Observations and Data Reduction}

\begin{figure*}
\includegraphics[scale=1.0]{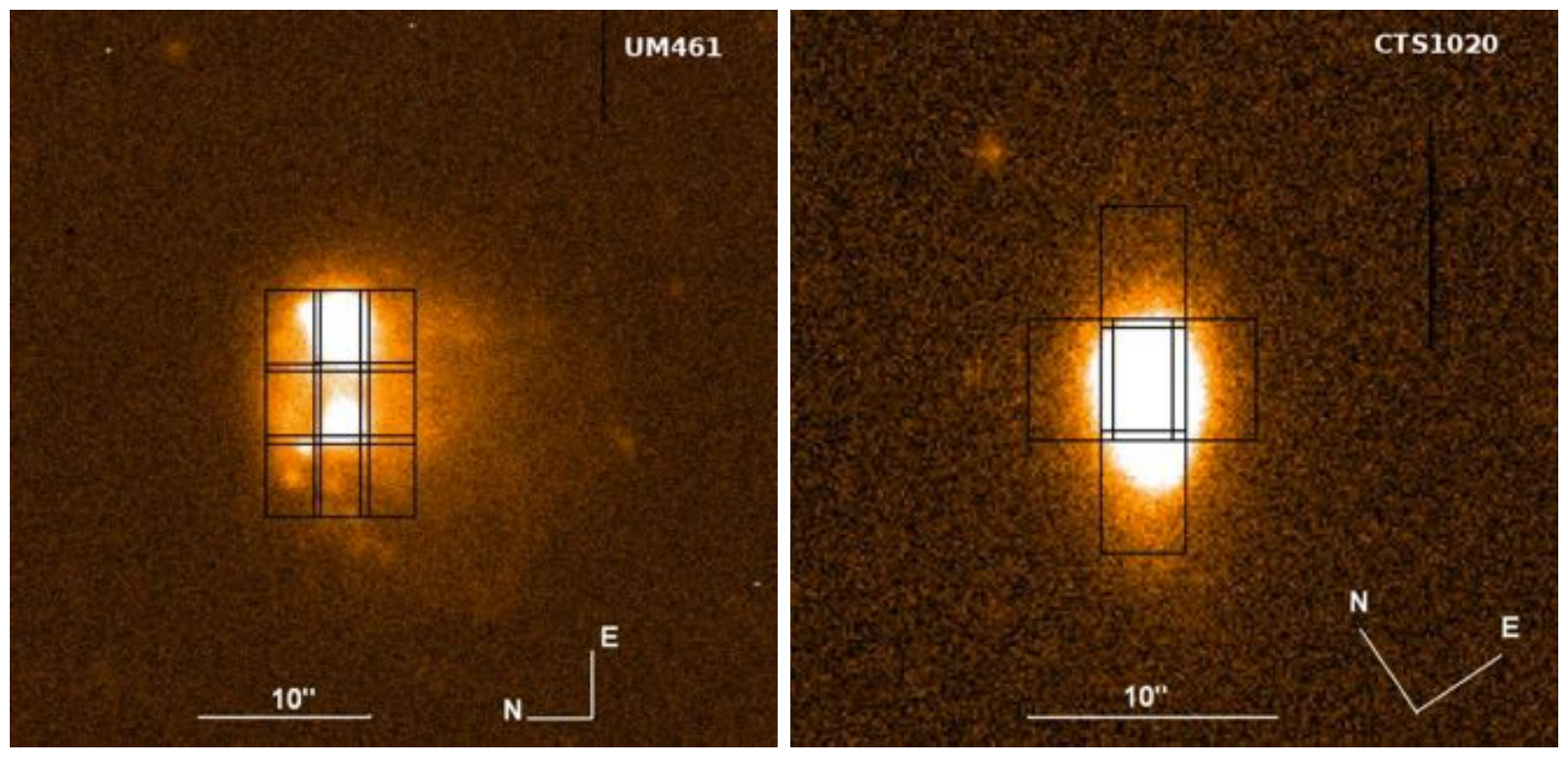}
\caption{Acquisition images of UM 461 and CTS 1020 taken with a $g$ filter and time exposure of 30s. Superimposed are the IFU field mosaic. Each IFU has a $3.5\arcsec \times 5\arcsec$ field of view.}
\label{Fig01}
\end{figure*}

Observations have been carried out in two runs  in February 12$^{th}$ 2008 and in February 18$^{th}$ 2010 as part of the GS-2008-Q32 program with Gemini South 8m telescope in Chile, using the Gemini Multi-Object Spectrograph (GMOS) equipped with an Integral Field Unit (IFU) on single-slit mode\citep{Allington2002}. Observations have been done under very good seeing conditions. UM 461 has been observed with a seeing between 0.35$\arcsec$ and 0.45$\arcsec$ and CTS1020 between 0.35$\arcsec$ and 0.5$\arcsec$.
Table~\ref{Table1} gives essentials characteristics for both galaxies. 
This mode uses 750 hexagonal lenses, each associated with a fiber, to sample the focal plane. 
The fibers are arranged to form single column positioned at the entrance of the slit of the spectrograph. The sky is sampled by 250 lenses at a distance of 1\arcmin\ from the science field. 
As individual IFU provides a field of view of $5\arcsec\times 3.5\arcsec$ (in single-slit mode), it was necessary nine and five fields to cover UM 461 and CTS 1020, respectively (see Fig.\ \ref{Fig01}). 
The time exposure of each field was 600s and the R831/550 grating has been used along with the g filter G5322 (see Table \ref{Table2} for the journal of observation). 
The IFU fields positioning are shown in Fig.\ \ref{Fig01} superimposed to the acquisition image taken with a $g$ filter and time exposure of 30 s.

\begin{table}
	\centering
	\caption{General parameters of the galaxies.}
	\label{tab:example_table}
	\begin{threeparttable}
	\makebox[\linewidth][r]{
	\begin{tabular}{lcccc}
		\hline
		  Galaxy & $\alpha$ & $\delta$  & $v_{sys}$ & $D$ \\ 
		  & $(J2000)$ & $(J2000)$  & (km s$^{-1}$) & (Mpc) \\
		\hline 
		UM 461   & 11 51 33.1 & -02 22 22 & 1039 & 19.3 \\
		CTS 1020 & 10 47 44.3 & -20 57 49 & 3789 & 57.4 \\
		\hline
	\end{tabular}
	}
        \begin{tablenotes}
            \footnotesize
            \item Informations obtained from NED.
        \end{tablenotes}	
	\end{threeparttable}
\end{table}

\begin{table*}
 \centering
\begin{minipage}{140mm}
  \caption{Journal of Observation}
\label{Table2}
  \begin{tabular}{ccccccc}
  \hline
   Galaxy     &   Observation      &   Instrument &  Exposure  & Fields  &  Grating &  Resolution \\
   		 &    Date 			&                      &   Time        &  Number  & Filter  &                      \\ \hline 
UM 461 	&     Feb. 12$^{th}$ 2008    &  GMOS/IFU  &  600s &  9  &  R831 - G5322  &    4396                  \\  
CTS 1020	&    Feb. 18$^{th}$ 2010  &  GMOS/IFU  &  600s  &  5  &   R831 - G5322   & 4396                 \\  
\hline
\end{tabular}
\end{minipage} \\
\begin{tablenotes}
\item Informations from Gemini website.
\end{tablenotes}
\end{table*}

Data have been reduced using special reduction package given by Gemini Staff \footnote{http://www.gemini.edu/node/10795} using the IRAF reduction package \footnote{Image Reduction and Analysis Facility is a software developed by the National Optical Astronomy Observatory - iraf.noao.edu}. All raw images have been bias corrected, trimmed and flat fielded. Flatfield images have also been used to locate the positions of the 750 lenses on the frame. Twilight images were used to estimate the grating response.
Arcs, from the CuAr lamp, have also been taken for the wavelength calibration. Using the bright O\,{\sc i} night sky line at 5577.338\,\AA\ we have estimated the wavelength accuracy to 0.1\,\AA. The last step of the reduction was the sky subtraction using the field located at 1\arcmin\ from the science field. Finally the data cube has been created with the {\sc gfcube}. We used a spacial resampling of 0.1\arcsec\ per pixel. 
The total spectral coverage for both objects is between 4442\,\AA\ and 6559\,\AA. From the arcs spectra, we deduced a spectral sampling of 0.33\,\AA\ per spectral pixel. 

\section{Ionized gas moment maps} \label{sec:IFUresults}

\begin{figure}
\vspace{-0.0cm}
\includegraphics[scale=0.22]{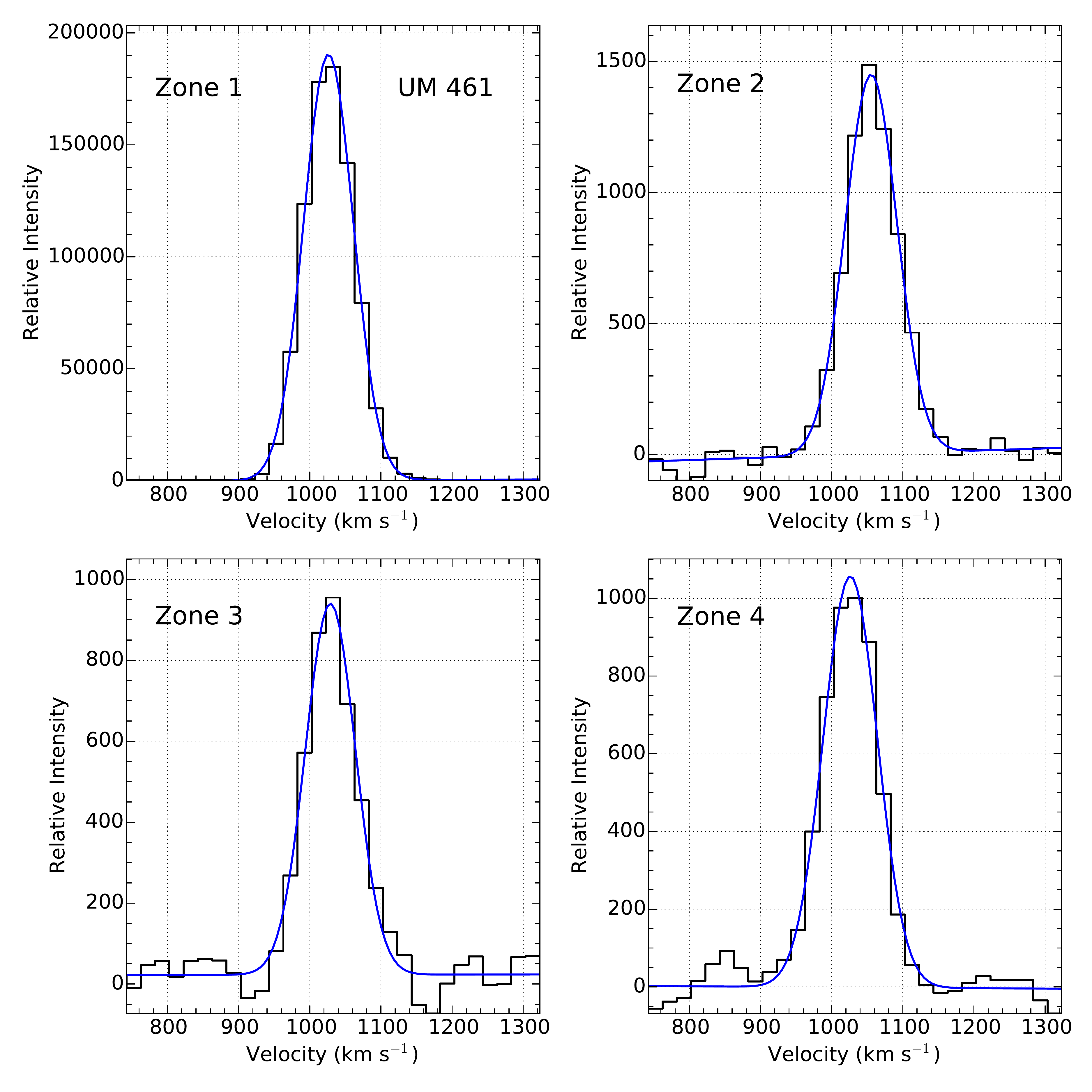}
\caption{UM 461: Profiles and Gaussian fit in four different zones of $0.4\arcsec~ \times~ 0.4\arcsec$. Different zones locations are given in Fig.\ \ref{Fig04}$b$. They represent integrated profiles in areas with very high SNR (>200) to area with SNR < 20.}
\label{Fig02}
\end{figure}

\begin{figure}
\vspace{-0.0cm}
\includegraphics[scale=0.22]{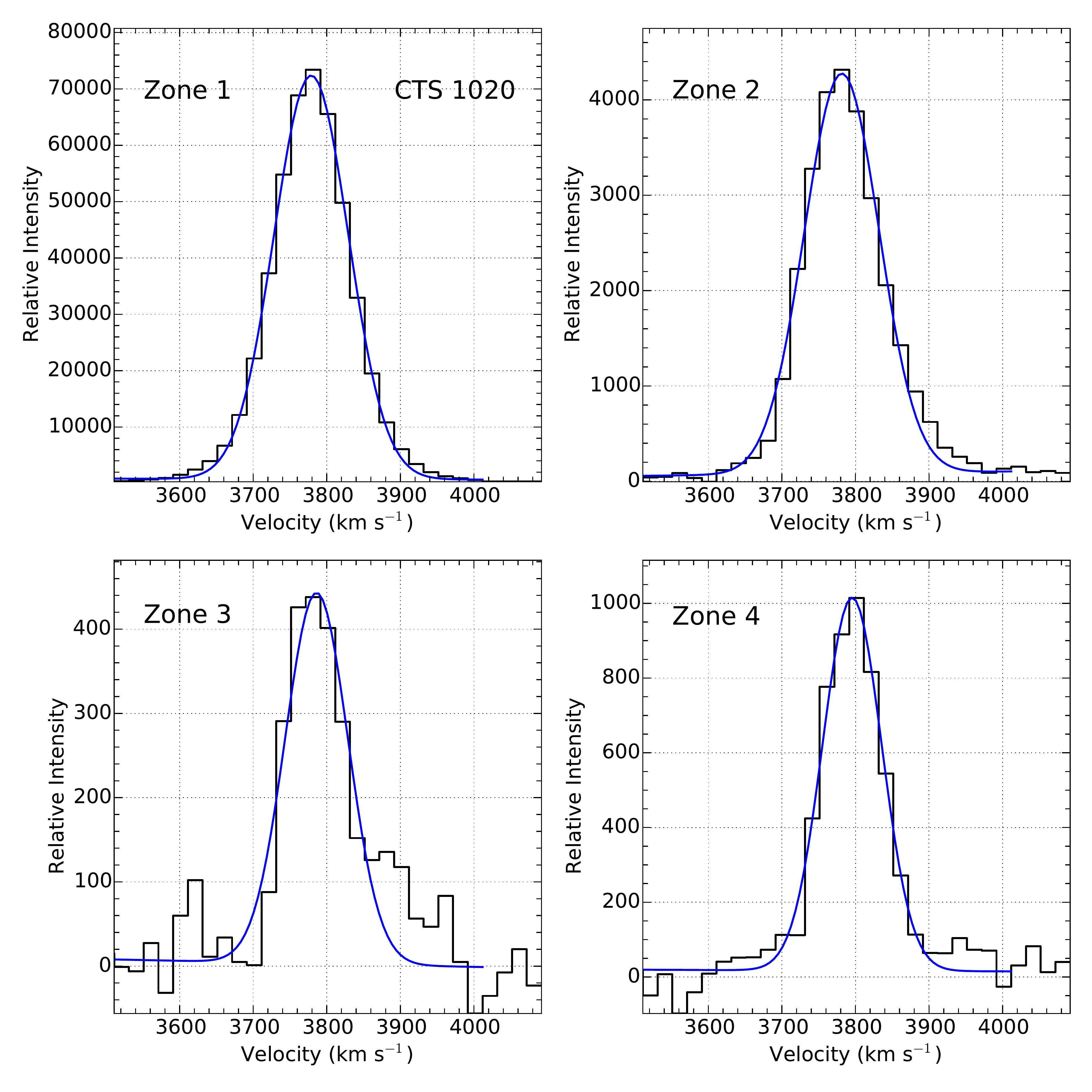}
\caption{CTS 1020: Profiles and Gaussian fit in four different zones of $0.4\arcsec~ \times~ 0.4\arcsec$. Different zones locations are given in figure Fig.\ \ref{Fig05}$b$. They represent integrated profiles in areas with very high SNR (> 200) to area with SNR < 20.}
\label{Fig03}
\end{figure}

\begin{figure*}
\includegraphics[width=5.8cm, angle=0, scale=1]{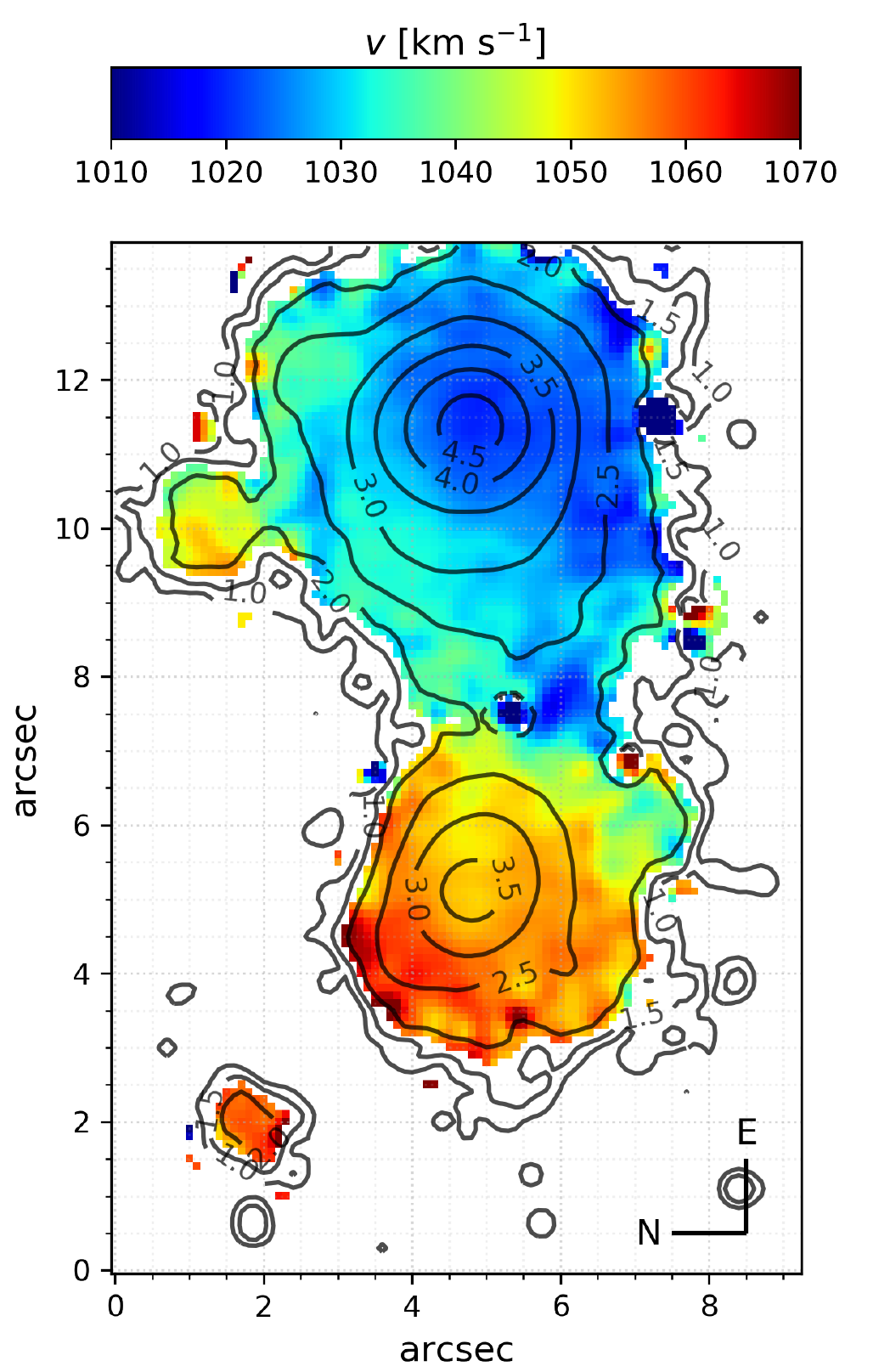}
\includegraphics[width=5.8cm, angle=0, scale=1.02]{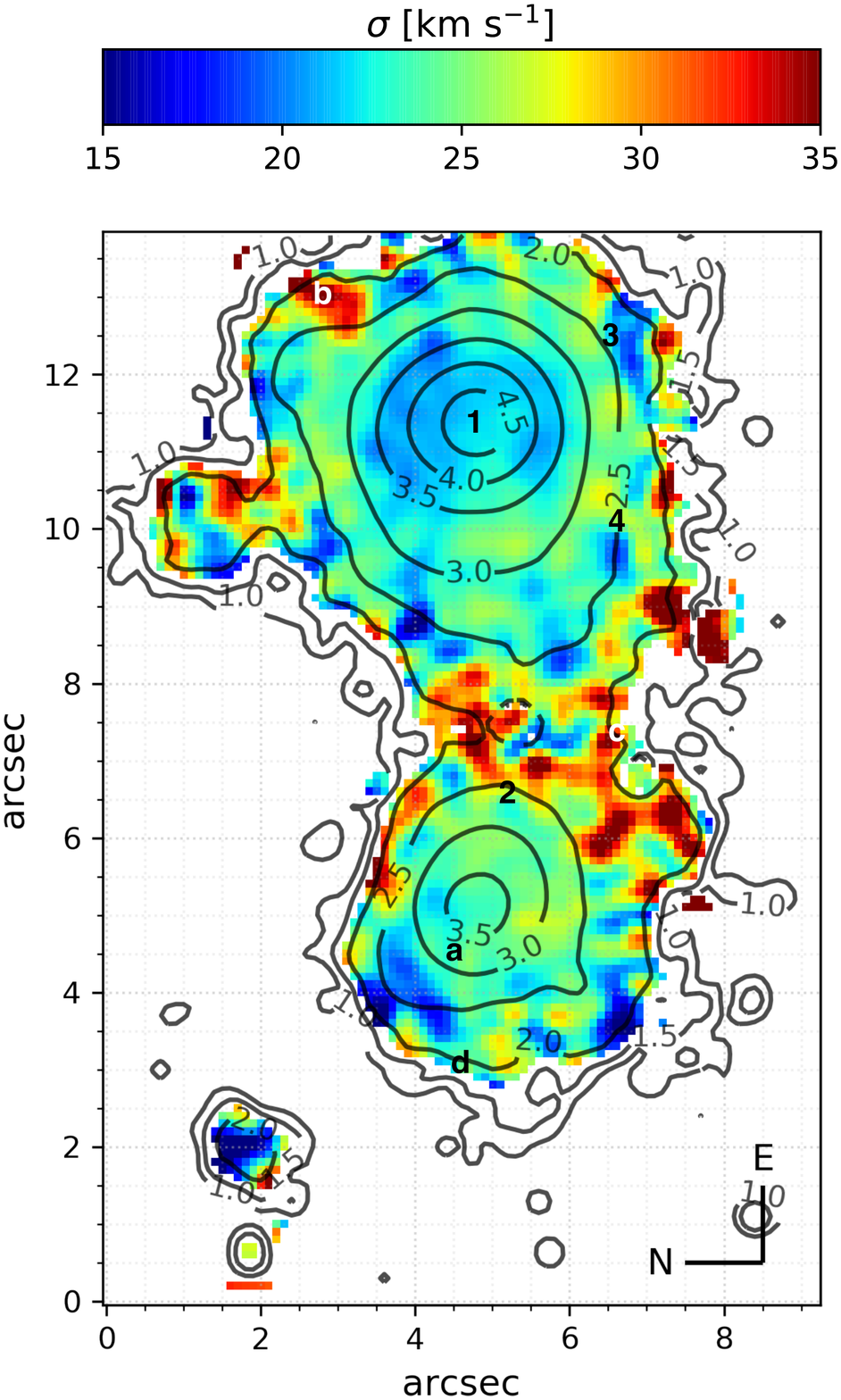}
\includegraphics[width=5.8cm, angle=0, scale=1]{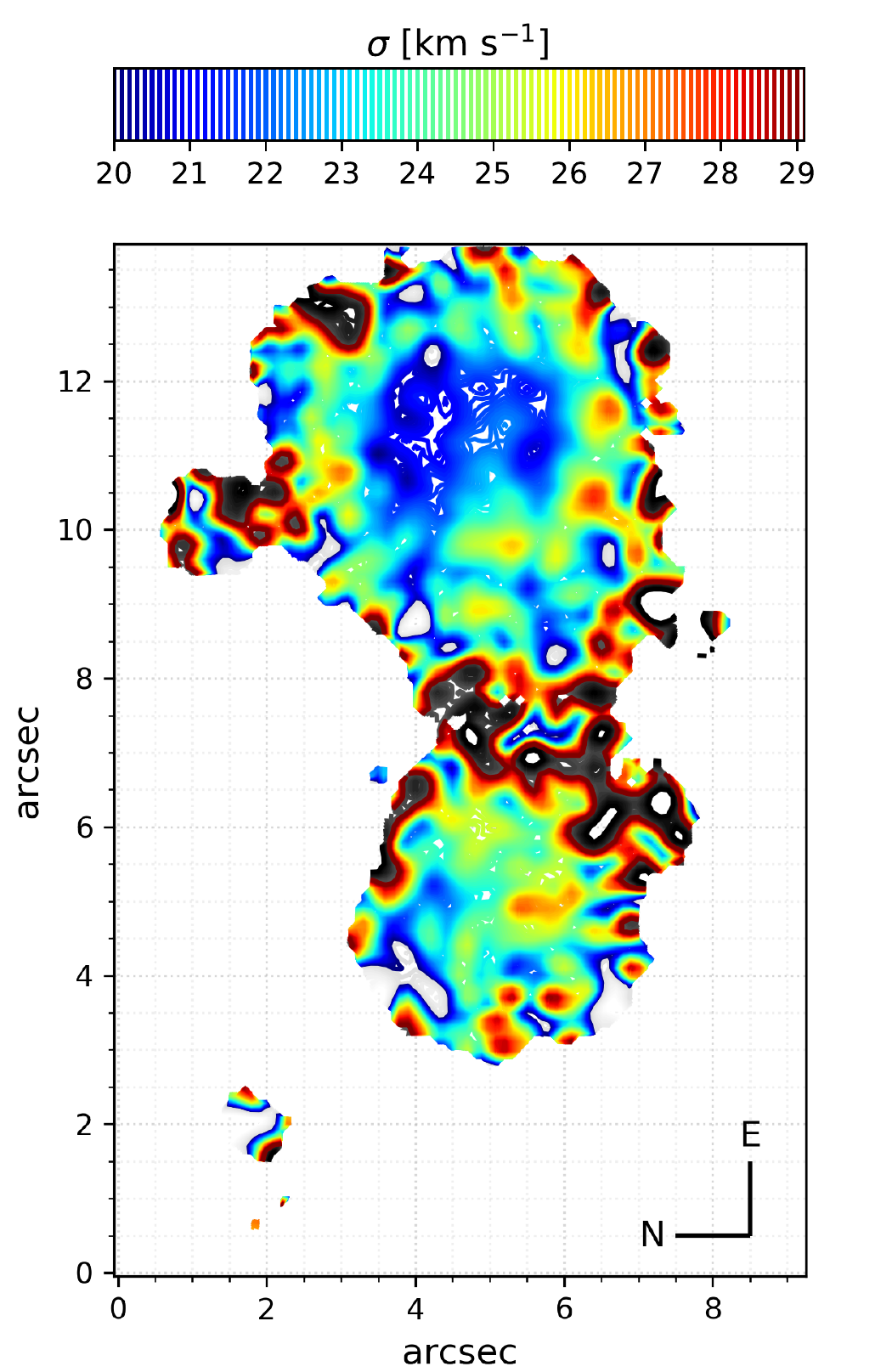}
\caption{UM 461: $a)$ Radial velocity and $b)$ velocity dispersion maps. The superimposed contours are the flux from [O\,{\sc iii}]$\lambda$5007 line. Labels refer extracted profiles in Fig.\ \ref{Fig02} (numbers) and Fig.\ \ref{Fig14} (letters). $c)$ Velocity dispersion contour map with isocurves varying by 0.1 km s$^{-1}$. In colors we have points in the interval 20--29 km s$^{-1}$ and in gray scale for values out of this interval.}
\label{Fig04}
\end{figure*}

Specific macro using IDL (Image Data Language)  have been used to produce the different maps.  We perform a linear fit of the continuum and a Gaussian fit of the emission lines, using the {\sc gaussfit} task in IDL.  We have considered signal from the galaxy a profile 3$\sigma$ above the sky level. The Gaussian fit gave us directly three maps: monochromatic (area of the profile), radial velocity (central wavelength) and velocity dispersion $(\sigma)$. 
Fig.\ \ref{Fig02} and Fig.\ \ref{Fig03} show profiles and the associated fit, estimated in 0.4$\arcsec$ x 0.4$\arcsec$ boxes, in four zones located in different areas for both objects. Location of zones is shown in Fig.\ \ref{Fig04}$b$ and Fig.\ \ref{Fig05}$b$. The Gaussian fit appears to be very good, even in outskirts zones where the SNR is lower ($\approx20$). Following the GAUSSFIT routine help, we estimate that the 1-$\sigma$ fit error is between 0.2 to 0.5 km s$^{-1}$ depending on SNR (see Fig.\ \ref{Fig02} and Fig.\ \ref{Fig03} captions).

For both galaxies, we were able to derive monochromatic map, velocity and velocity dispersion maps using only [O\,{\sc iii}]$\lambda$5007 emission line. 
The velocity dispersion map has been corrected from the instrumental and thermal broadening.  
The velocity dispersion ($\sigma$) was obtained by correcting the observed one ($\sigma_ {obs}$) for both instrumental ($\sigma_{inst}$) and thermal ($\sigma_{th}$) broadening, $\sigma_{obs} =\sqrt{\sigma^2 + \sigma^2_{th} + \sigma^2_{inst}}$.
We have estimated the thermal dispersion to $\sigma_{th} = 3.2 \mathrm{\ km\ s}^{-1}$ for the oxygen (at a temperature of $10^4K$), and  $\sigma_{inst} = 27.5 \pm 0.1$ km s$^{-1}$ for both galaxies, from several emission lines in calibration lamp used for the wavelength calibration. 
To take into account the seeing quality of observations, we performed a Gaussian smoothing of 0.4$\arcsec$ FWHM of the maps.
Tests in regions on the outskirts of galaxies, where the SNR is below 20, reveal that sigma estimation from the gaussian fit has an relative error less than 3\% when continnum level is estimated till three times the linear fit errors.

\subsection{UM 461}

\subsubsection{Monochromatic map}

The [O\,{\sc iii}]$\lambda$5007 monochromatic emission map is superimposed on both velocity and velocity dispersion maps presented in Fig. \ref{Fig04}$a$ and Fig. \ref{Fig04}$b$, respectively. 
The eastern knot is a factor of 12 brighter compared to the western knot.

\subsubsection{Radial Velocity map}
The velocity map has been elaborated using a systemic velocity v$_{syst}$ = 1039 km s$^{-1}$ from NASA/IPAC Extragalactic Database (NED)\footnote{http://ned.ipac.caltech.edu/}.   
The eastern knot shows a velocity gradient of 15 km s$^{-1}$ from the southeast to the northwest of the knot.
The western knot has a different pattern, as does not exhibit a clear gradient and is dominated by velocities between 1048 and 1061 km s$^{-1}$.
Despite the low amplitude gradient in the eastern knot, the overall motion of the ionized gas is not ordered. 
Considering the stellar clusters and complexes spread along the galaxy, as observed in the near-infrared \citep{Lagos2011,Noeske2003}, the overall kinematics could result from the interaction of these stellar populations with the interstellar medium, as they evolve and inject mechanical energy into the medium through stellar winds and supernova explosions.
The eastern knot, despite more compact, has at least three smaller companions around it, which could be responsible for the disturbed velocity field, mainly in the northeastern region. In the western knot, the stellar clusters are more spread out among them, producing a much less ordered motion. In fact, the velocity map of UM 461 suggests a weak kinematical connection between the knots. In agreement with this scenario, \citet{OlmoGarcia2017} show evidence of stellar feedback in the eastern knot and its coherent motion within the host galaxy. \citet{Lagos2018} velocity field of UM 461 is consistent with ours, showing the same velocity gradient in both knots. Radial velocities show a difference of  $\approx 15$ km s$^{-1}$, that can be attributed to the spectral resolution difference.

\subsubsection{Velocity dispersion map}
The velocity dispersion map in Fig.~\ref{Fig04}$b$ shows that all velocity dispersions are supersonic, ranging between 15 to 35 km s$^{-1}$ in a few pixels.
In the eastern knot, it is noticeable a ring-like structure with $\sigma$ values between 25 and 29 km s$^{-1}$ that envelops a region of lower dispersion (21--24 km s$^{-1}$).
\citet{Melnick1988} and \citet{Bordalo2011} reported slightly lower velocity dispersion of 14.5 km s$^{-1}$ and 12.6 km s$^{-1}$, respectively. Our lowest $\sigma$ is 15 km s$^{-1}$, but in the center of the brightest knot, where the one fiber measurement certainly has been done, our
value is $\approx20$ km s$^{-1}$. 
In the western knot, the velocity dispersion increases toward the southeast, with some small regions of increasing $\sigma$.
A few pixels show very high velocity dispersion (especially in the connection between the knots) up to 50 km s$^{-1}$.  
In Fig.~\ref{Fig04}$c$ we show a velocity dispersion contour map obtained varying the isocurves by 0.1 km s$^{-1}$. In this map, we wanted to highlight regions in the interval 20 to 29 km s$^{-1}$ (shown in colors), because the contrast with the highest and lowest $\sigma$ values (in gray scale) makes it difficult to see some fine structures within these regions. In fact, we found several regions of increasing $\sigma$ (meaning a local $\sigma$ peak surrounded by decreasing values) along the galaxy. By restricting the $\sigma$ range allowed us to better distinguish the ring-like structure in the eastern knot. This structure resembles the distribution of H$\beta$ equivalent width reported by \citet{Lagos2007}, that exhibits $EW(H\beta$) higher than 300\,\AA, indicating an intense star formation activity in the eastern knot.

\begin{figure*}
\includegraphics[width=5.8cm, angle=0, scale=1.]{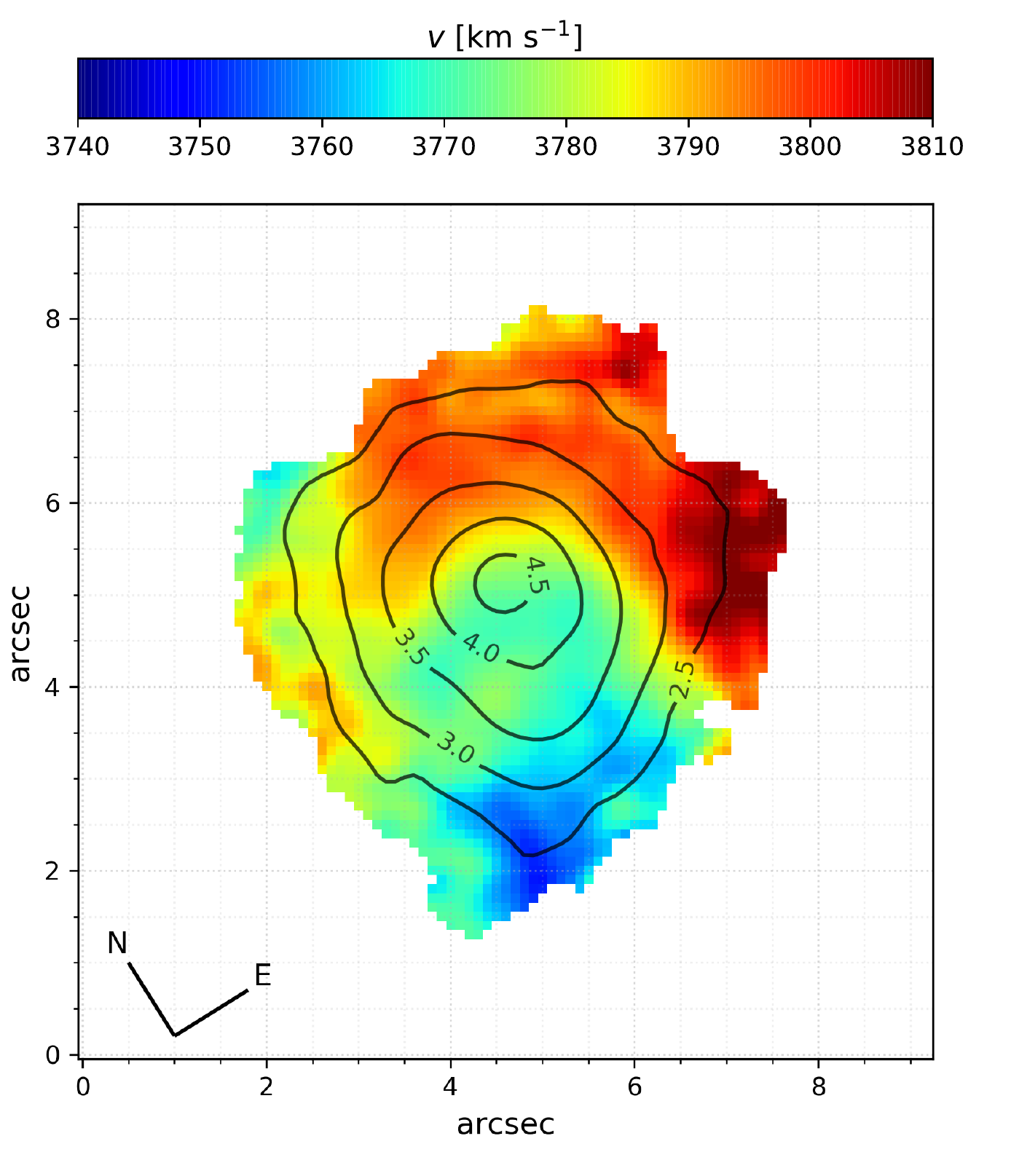}
\includegraphics[width=5.8cm, angle=0, scale=1.]{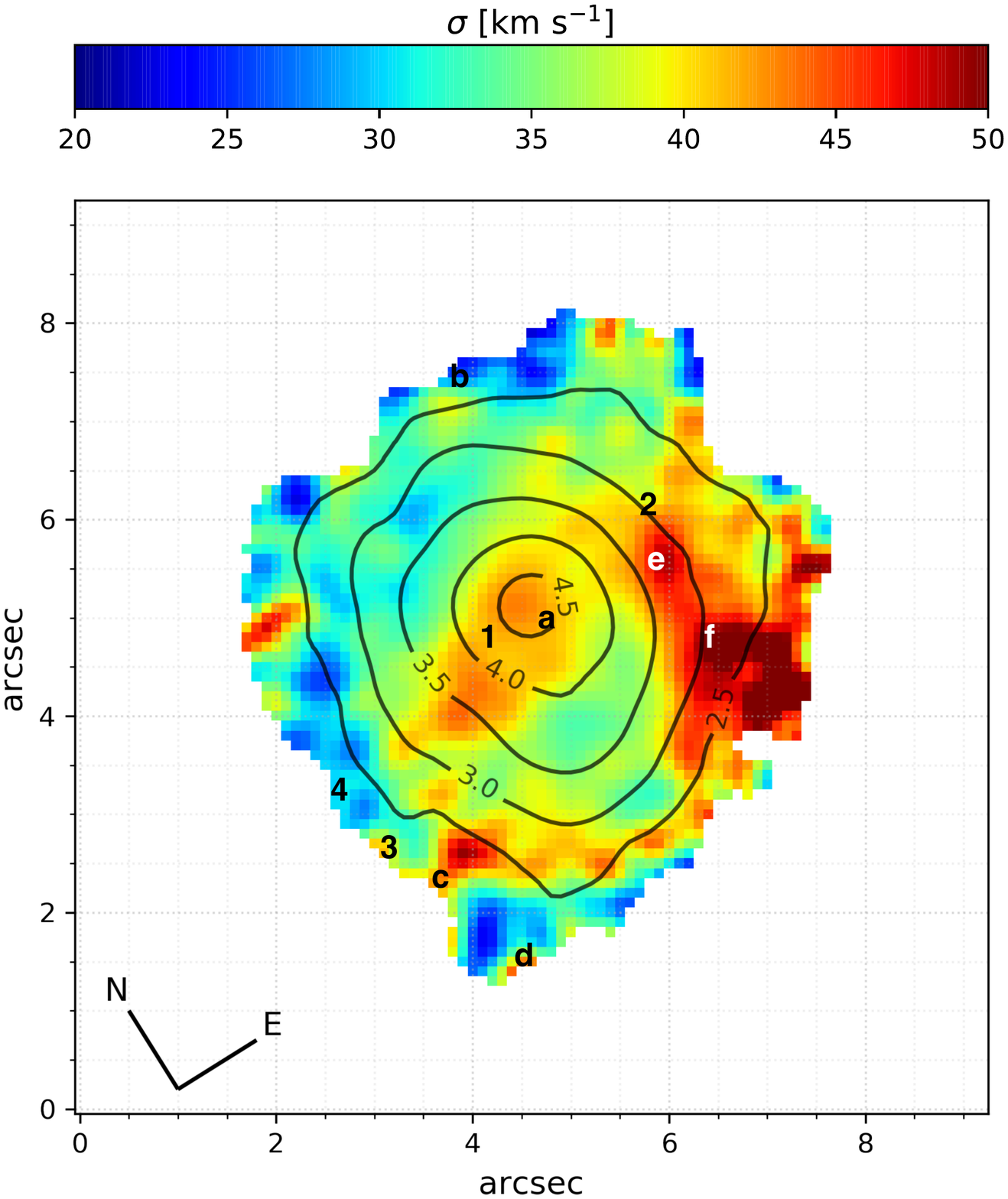}
\includegraphics[width=5.8cm, angle=0, scale=1.]{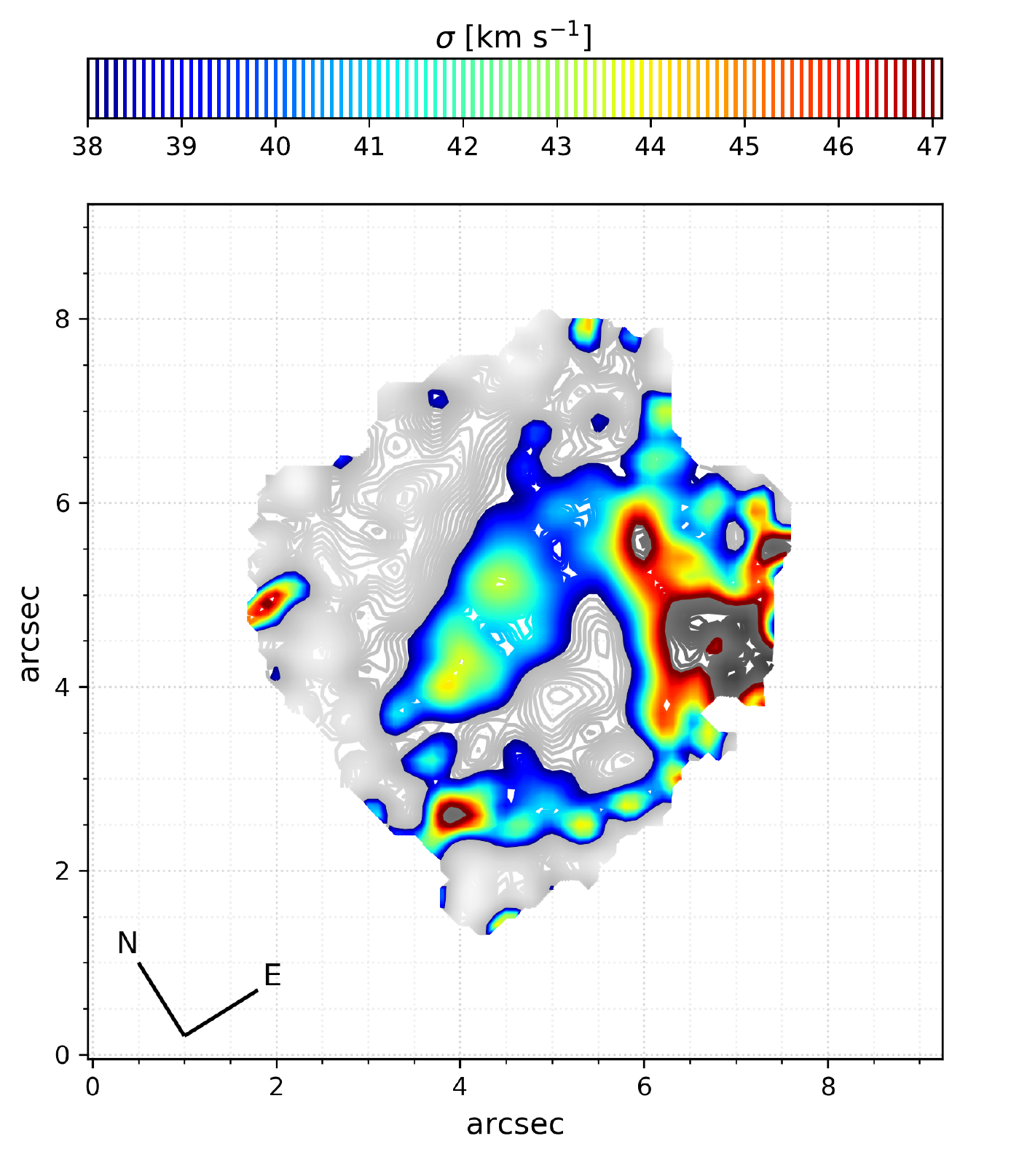}
\caption{CTS 1020: $a)$ Radial velocity and $b)$ velocity dispersion maps. The superimposed contours are the flux from [O\,{\sc iii}]$\lambda$5007 line. Labels refer extracted profiles in Fig.\ \ref{Fig05} (numbers) and Fig.\ \ref{Fig15} (letters). $c)$ Velocity dispersion contour map with isocurves varying by 0.1 km s$^{-1}$. In colors we have points in the interval 38--47 km s$^{-1}$ and in gray scale for values out of this interval.}
\label{Fig05}
\end{figure*}

\subsection{CTS 1020}

\subsubsection{Monochromatic map}
The [O\,{\sc iii}]$\lambda$5007 monochromatic emission map is superimposed on both velocity and velocity dispersion maps presented
in Fig.\ \ref{Fig05}$a$ and Fig.\ \ref{Fig05}$b$, respectively. 
Emission extends across all observed fields (and possibly beyond the observed fields in the northeast and east). It is also noticeable an extension towards the southwest, which probably is a manifestation of a second nucleus, but the lack of spatial resolution prohibits to see it.

\subsubsection{Radial Velocity map}
We used v$_{syst}$ = 3789 km s$^{-1}$ as systemic velocity from NED.
The velocity field of CTS 1020 is very different from UM 461. Here we can clearly see a velocity gradient more consistent with a 
rotating disk. Across a velocity major axis of 0$^o$, radial velocity varies from 3752 km s$^{-1}$ in the south to 3800 km s$^{-1}$ in the north. Southwest from the monochromatic center (at $\sim1\arcsec$), a small region shows radial
velocities 10 km s$^{-1}$ higher than immediate surroundings. We speculate that it can be a high velocity cloud. East of the center, on the edge of the field, a region of $2\arcsec \times 1.5\arcsec$ shows high radial velocities between 3805 and 3818 km s$^{-1}$, the highest of the velocity field. Here too, we can speculate that it is a high velocity cloud, but because our fields did not cover the galaxy entirely, it is difficult to conclude. Despite a clear velocity gradient compatible with a disk pattern, it is almost certain that the kinematics is more complex than it seems. 

From the monochromatic map we derived an inclination angle of the disk by doing the ratio of the minor axis to the major axis, we found 30$^o$. 

\subsubsection{Velocity dispersion map}
The outskirt of the velocity dispersion map (Fig.\ \ref{Fig05}$b$) shows velocity dispersion lower than $\sigma = 20$ km s$^{-1}$, where the SNR is
close to the limit discussed above and the Gaussian fit profile is not so good. Nevertheless, after an eye check of these profiles, we are confident
that the estimated dispersion is accurate. 
Beside this region, the velocity dispersion reaches low values ($\sigma<38$ km s$^{-1}$) in the northern and southern regions, while highest values ($\sigma>47$ km s$^{-1}$) are concentrated in an arch in the southeastern part of the galaxy. Crossing the center of the galaxy in the east to west direction $\sigma$ assumes values between 38 and 47 km s$^{-1}$.
Fig.\ \ref{Fig05}$c$ shows the $\sigma$ contour map restricted to the range 38--47 km s$^{-1}$ (shown in colors). These values are mainly concentrated in the area along the center toward the west of the galaxy.
By restraining the $\sigma$ range, we can now distinguish few regions of increasing $\sigma$: in the center, $\sigma$ increases with intensity; 
in the west coincides with the disturbed region seen in the velocity field; in the southeastern region $\sigma$ reaches values higher than 47 km s$^{-1}$.

\section{Diagnostics diagrams}\label{sec:diagrams}

\subsection{Description}
\begin{figure*}
\includegraphics[width=17.0cm, angle=0, scale=1.1]{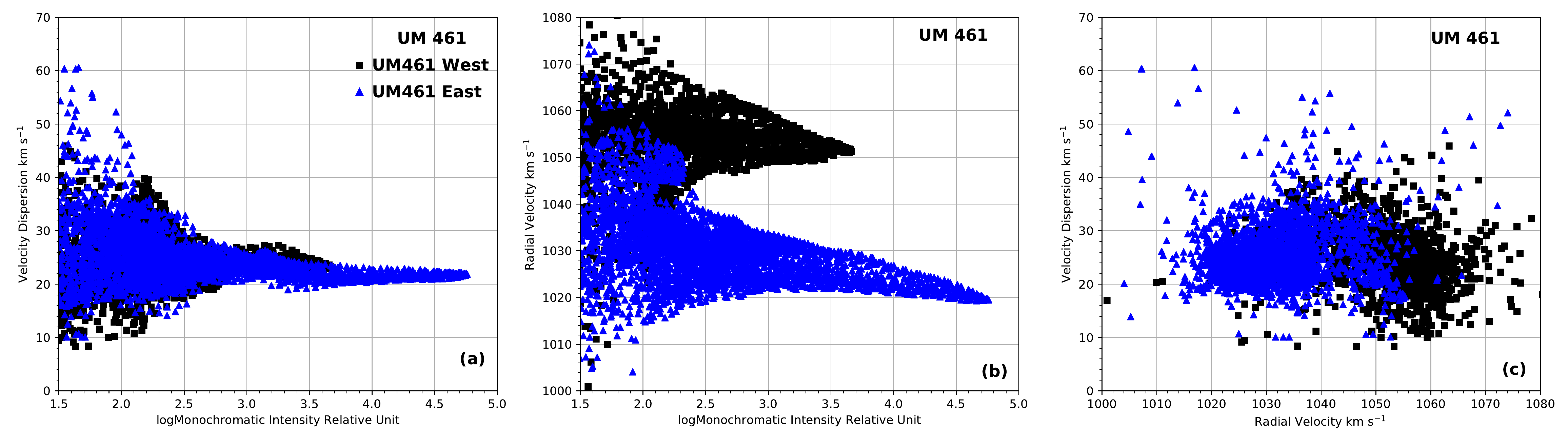}
\includegraphics[width=17.0cm, angle=0, scale=1.1]{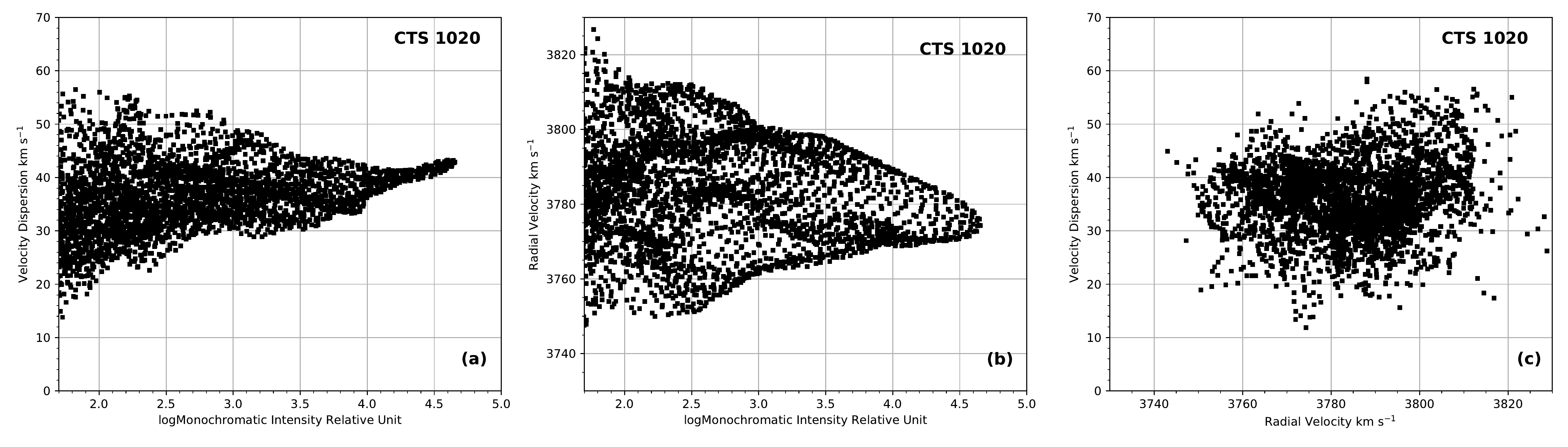}
\caption{Diagnostic diagrams for UM 461 ({\it top}) and CTS 1020 ({\it bottom}): (a) Velocity Dispersion vs Intensity; (b) Radial Velocity vs Intensity; and (c) Velocity Dispersion vs Radial Velocity (c). We have separated the western (black) and eastern (blue) knots for UM 461. Horizontal dashed line represents instrumental broadening.}
\label{Fig06}
\end{figure*}

As mentioned before, several studies have shown that \hbox{H\,{\sc ii}} galaxies have supersonic velocity dispersion and little velocity gradient. In order to study the dynamic of these objects, several studies have suggested over the years, the use of the so called, diagnostic diagrams such: $I - \sigma$, $I - V_r$ or $V_r - \sigma$. For each galaxy, we are presenting the $I - \sigma$ in Fig.~\ref{Fig06}$a$, $I - V_r$ in Fig.~\ref{Fig06}$b$ and $V_r - \sigma$ in Fig.~\ref{Fig06}$c$. \citet{Bordalo2009} summarise the different interpretations  of these diagrams. Introduced by \citet{MunozTunon1996}, the $I - \sigma$ diagram has been used by those authors to identify expanding shells by localizing inclined bands. This interpretation is based on the fact that the velocity dispersion should be higher at the center of the shell and the intensity lower because less material is crossed along the line of sight than at the inner and outer edges of the shell. Large variations of radial velocity with a relatively narrow intensity interval in a $I - V_r$ diagram, could mean an expansion or an inflow of matter.  The $V_r - \sigma$ panel looks at the dependence between the two variables. If a significative correlation is found between velocity and dispersion, it could mean the presence of relative motion inside the system. The inclined pattern is the signature of systematic motion, such as Champagne flows such that cloud of gas with high $\sigma$ moves away from us (positive slope) or toward us (negative slope).

\subsubsection{UM 461}
The top panels of Fig.~\ref{Fig06} shows for UM 461. Each knot is represented with two different symbols. The  $I - \sigma$ plot shows a similar behaviour for both knots, a trumpet like shape where the low intensity region has a large velocity dispersion range and the high intensity a narrow velocity dispersion interval. 
The  $I - V_r$ panel clearly shows the velocity differences between both knots, with the western knot moving away from the eastern knot. The radial velocity range is larger at very low intensity level, due to the lowest SNR. No vertical bands, characteristic of expansion motion, is visible in either knots. The  $V_r - \sigma$ plot shows a clear correlation, but a closer look seems to show sub-structures. We decided to investigate further using robust statistical tools, detailed in Section \ref{sec:statistics}.

\subsubsection{CTS 1020}
The bottom panels of Fig.~\ref{Fig06} show the diagrams for CTS 1020. 
The $I-\sigma$ diagram exhibits a different behaviour of $\sigma$ compared to UM 461. Here, $\sigma$ increases with intensity until reaches a lane along all intensities with values between 38 and 43 km s$^{-1}$ (highlighted in Fig. \ref{Fig05}$c$), and a mean of $40.3\ \mathrm{km\ s^{-1}}$. 
Fig.~\ref{Fig06}$b$ shows the $I - V_r$ corresponding diagram for this object. No radial motions (materialised by a vertical band) can be seen. The triangular shape toward the higher intensity, is more characteristic of a rotation pattern, with a maximum velocity amplitude of 80 km s$^{-1}$. The third plot (Fig.~\ref{Fig04}$c$) represents the $V_r - \sigma$ diagram. It seems that two populations are present, we have performed a deep statistical analysis in the Section \ref{sec:statistics} in order to extract those two populations. 

\subsection{$I - \sigma$ diagram}
In Figs.\ \ref{Fig07} and \ref{Fig08}, we further explored the $I-\sigma$ diagram by looking at different regions and their respective location in the velocity dispersion map.
We have divided the $I-\sigma$ diagram in three intervals of intensity (\it high \rm -- $\log I>3.5$; \it intermediate \rm -- $2.7<\log I<3.5$; and \it low \rm -- $\log I<2.7$), and $\sigma$ intervals according to values in the horizontal lane, in order to distinguish their distribution over the galaxy. 

\subsubsection{UM 461}

\begin{figure*}
\centering
\includegraphics[width=17cm, angle=0, scale=1.11]{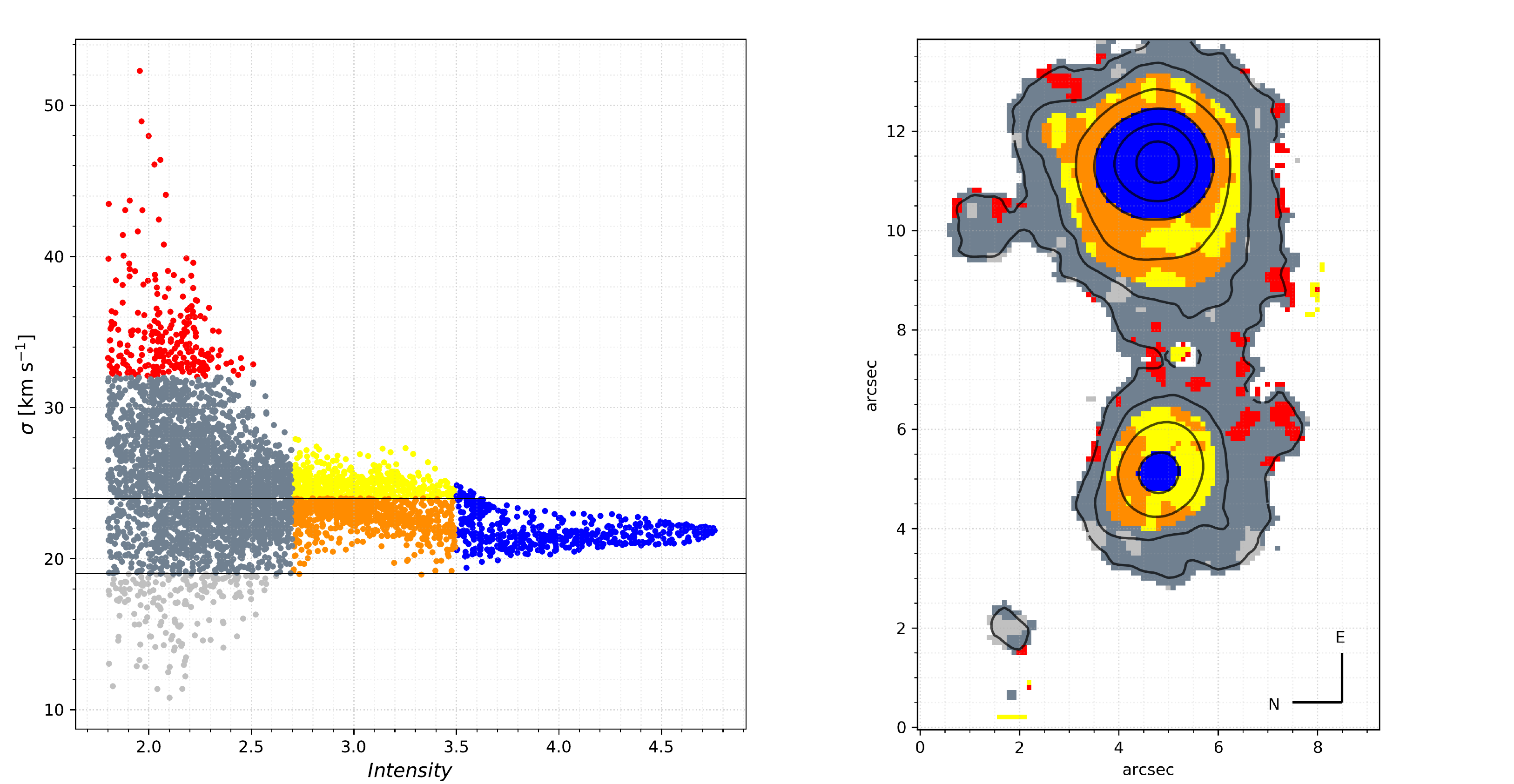}
\caption{$I-\sigma$ diagram and respective velocity dispersion map for UM 461. The black horizontal lines indicate, from top, the velocity dispersion in 24 and 19 km s$^{-1}$. The intensity is given in arbitrary units and is shown as contours in the map.}
\label{Fig07}
\end{figure*}

Fig.\ \ref{Fig07} shows the $I-\sigma$ diagram and the respective velocity dispersion map for UM 461. 
This galaxy has a simple morphology with intensity decreasing outwards the knots. The high intensity regions (in dark blue) have an almost constant velocity dispersion ($\langle\sigma\rangle\sim21.8\ \mathrm{km\ s^{-1}}$), which also comprises an horizontal lane along the intensity range. The regions in $2.7<\log I<3.5$ also exhibit values around $\langle\sigma\rangle$ (in orange -- mainly in the eastern knot) with some regions of higher $\sigma$ (in yellow), which coincides with the ring-like structure shown in Fig.\ \ref{Fig04}$c$. 
The lowest ($\sigma<20$ km s$^{-1}$) and highest $\sigma$ values ($\sigma>32$ km s$^{-1}$) are located in the outermost regions of the galaxy, surrounding both star-forming knots.
Only a few pixels reach $\sigma$ values higher than 32 km s$^{-1}$.
These regions (shown in grey and red) cover the whole $\sigma$ range and form a triangular pattern, which is related to the turbulent motion in the diffuse gas that permeates the star-forming regions in the galaxy \citep{Moiseev2012,Moiseev2015}. 
It was not possible to identify signatures of expanding shells, as proposed by \citet{MunozTunon1996}.
The horizontal lane 
with an almost constant $\sigma$ is supposed to be a supersonic random motion caused by a constant passage of bow shocks from low-mass stars in the model of \citet{MunozTunon1996}.

\subsubsection{CTS 1020}

\begin{figure*}
\includegraphics[width=17cm, angle=0, scale=1.]{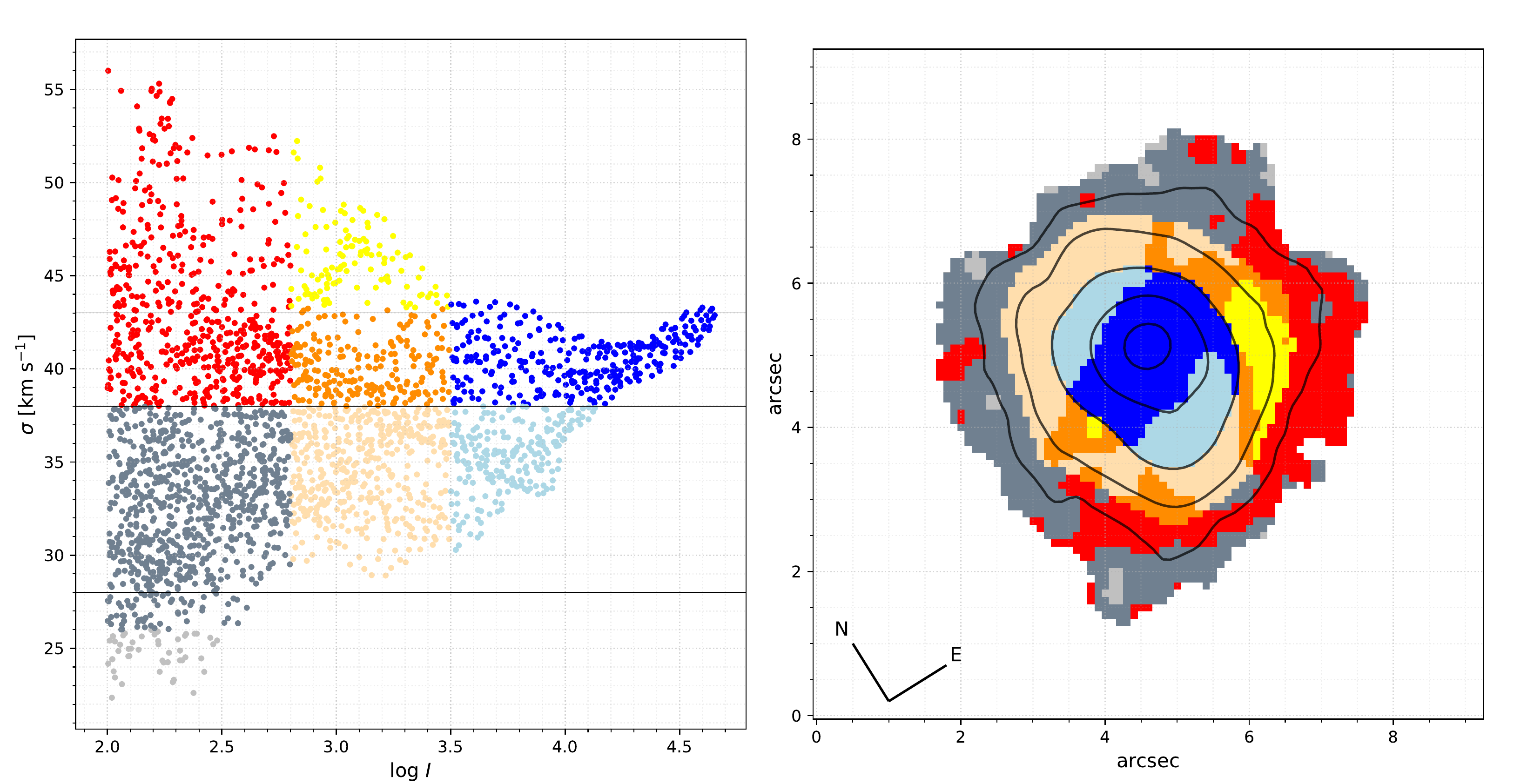}
\caption{$I-\sigma$ diagram and respective velocity dispersion map for CTS 1020. The black horizontal lines indicate, from top, the velocity dispersion in 43, 38 and 28 km s$^{-1}$. The intensity is given in arbitrary units and is shown as contours in the map.}
\label{Fig08}
\end{figure*}

In CTS 1020, the velocity dispersion distribution (Fig. \ref{Fig08}) shows a different pattern from that of UM 461.
Although the $\sigma$ range also decreases with intensity, the regions of high intensity are related to high velocity dispersion.
This galaxy also presents a horizontal lane of an almost constant velocity dispersion ($\langle\sigma\rangle\sim40.3$ km s$^{-1}$), but contrary to UM 461, in CTS 1020 this value is higher than the overall mean ($37.1$ km s$^{-1}$) along the galaxy.
The points in this lane (represented in dark blue, dark orange and red) are not widely spread along the galaxy (see also Fig.\ \ref{Fig05}$c$), as seems to be the case of UM 461. Instead, the regions represented in light blue, light orange and light grey, seems to cover a large area of the galaxy. 
The velocity dispersion seems to decrease outwards, but some regions show a velocity dispersion higher than the local average, such as the regions in dark orange. 
The high intensity region with $\sigma$ values around $\langle\sigma\rangle$ (in dark blue) cross the center of the galaxy in the east to west direction (As also seen in Fig.\ \ref{Fig05}$c$.). 
As observed in UM 461, the lowest ($\sigma<28$ km s$^{-1}$) and highest $\sigma$ values ($\sigma>43$ km s$^{-1}$) are located in the outer parts of the galaxy,
except by a small region with $\sigma$ higher than 43 km s$^{-1}$ in the southeast (in yellow), which would be an evidence of expanding motion. 
Despite the different pattern, the main similarity with UM 461 is that the outermost regions (shown in grey and red) cover the whole $\sigma$ range, and could also be related to the turbulent motion of the diffuse gas.

\section{Statistical Analysis}\label{sec:statistics}

In order to extract information from the $V_r - \sigma$ diagram, we performed statistical analysis to determine the possibility of having several independent populations.
To achieve this goal, we employed the {\tt R} statistical package ({\tt R} Development Core Team 2009), largely used in different statistical analysis. 
We aim at finding how many independent components are present (task {\tt Mclust}), to locate them in the diagram and in the map (so-called geographic location).
{\tt Mclust} is a {\tt R} function for model-based clustering, classification, and density estimation based on finite Gaussian mixture modeling. 
An integrated approach to finite mixture models is provided with routines that combine model-based hierarchical clustering and several tools for model selection \citep[see][]{Fraley2007}.

A central question in finite mixture modelling is how many components should be included in the mixture. In the multivariate setting, the volume, shape, and orientation of the covariances define different models (or parametrisation) with their different geometric characteristics.
In {\tt Mclust}, the number of mixing components and the best covariance parameterisation are  selected using the Bayesian Information Criterion (BIC).
{\tt Mclust} also relates each  element in the dataset to a particular component in the mixture. 

The code uses the Expectation-Maximization (EM) algorithm that maximizes the conditional expected log-likelihood at each M-step of the algorithm. 

\begin{figure*}
\vspace{-0.5cm}
\includegraphics[width = 16.0cm]{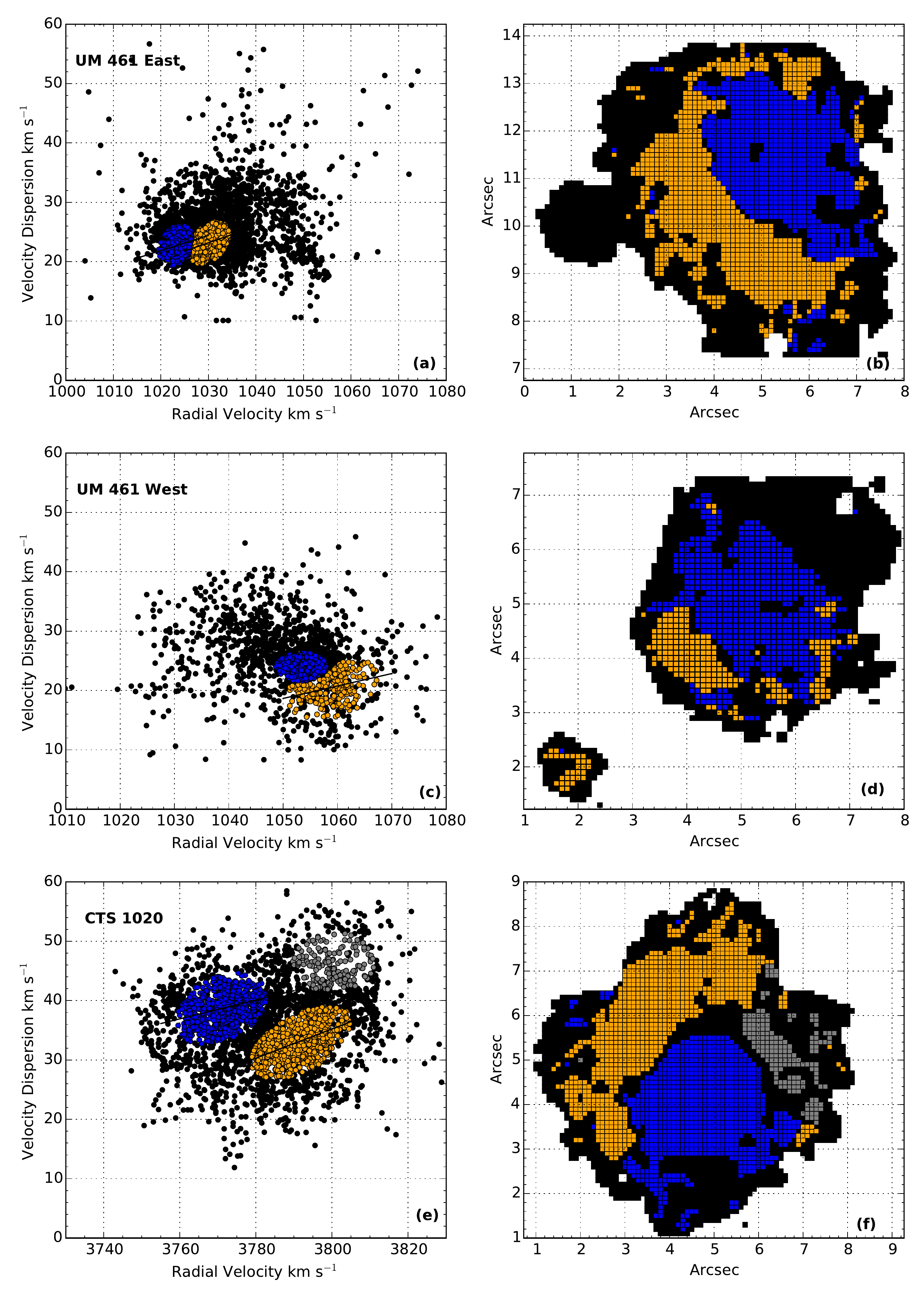}

\caption{UM 461 East (a) $V_r - \sigma$ diagram. Two components, blue and orange, separated using statistical analysis with the task {\tt Mclust}. (b) Geographic representation of the different components. UM 461 West (c)  $V_r - \sigma$ diagram. {\tt Mclust} decomposition result, (d) Geographic representation of the different components.  CTS 1020 (e) $V_r - \sigma$ diagram. Five components separated using {\tt MClust} task. CTS 1020 (f) Geographic representation of the different components. Solid lines represent linear regression of the different components in all plots.}
\label{Fig09}
\end{figure*}

Below, we are detailing how we applied these different tasks and their results.

\subsection{UM 461}
In this case, we had to separate knots on the East and West. Fig.~\ref{Fig09}$a,~b$ shows the result of such statistical analysis for the eastern knot and Fig.~\ref{Fig09}$c,~d$ western knot. 

In UM 461 eastern knot, we found  $m~=~4$ components. Two of these components did not seem to have physical meaning since they were regrouping dispersed points in the galaxy outskirt and small areas. The two others components correspond to two peaks in a density map of the diagram (not presented here), which gave us more confidence in the statistical decomposition.

Both components are plotted in Fig.~\ref{Fig09}$a$, respectively in blue and orange, within a 80\% confidence level. We perform a standard Pearson's product-moment correlation test for the different components, in order to show the existence of systematic motions mentioned before. Subcomponent 1 (in orange) has a correlation of $0.37$ and correlation of subcomponent 2 (in blue) is $0.28$, both are considered as weak correlation.  

In UM 461 western knot, we also found $m~=~4$ components. As previously, only two seems to show a physical meaning. The relevant components correspond to denser areas of the diagram. The Pearson test, in both components gives respectively $0.33$ (orange component) and $-0.01$ (blue component). 

In both, East and West regions, we performed a simple linear regression (showed as solid lines in Fig.~\ref{Fig09}$a$, Fig.~\ref{Fig09}$c$) for these components when the correlation is relevant. In this context, both subcomponents in the eastern region can be interpreted as complexes with relatively high dispersion, moving away from the observer (positive slope). In the western region only the orange component shows a relevant correlation and the 
linear regression has a positive slope, corresponding to a complex moving away from the observer.

Fig.~\ref{Fig09}$b,~d$ show the geographic location of the different components and subcomponents for both eastern and western knots. 

In both knots, the geographic location is compatible with the velocity field represented in Fig.~\ref{Fig04}$a$, with the subcomponent in blue representing the low velocity area and the orange subcomponent representing higher velocities.

\subsection{CTS 1020}

As for UM 461, before analysis, $V_r$ and $\sigma$ were normalized in order to avoid bias due to amplitude differences between $V_r$ and $\sigma$. 

The {\tt Mclust} analysis is presented in Fig.~\ref{Fig09}$e,~f$. 
It has separated three components in the $V_r - \sigma$ diagram. All of  three components appear to have a physical meaning when we look at the X-Y location map (Fig.~\ref{Fig09}$f$).  The first two ones (blue and grey) show weak (0.3 for first component) and moderate (0.54 for the second) correlation. 

Fig.~\ref{Fig09}$f$, representing the X-Y location of these regions, is also compatible with CTS 1020 velocity map (Fig.~\ref{Fig05}$c$) where the orange region represents the high radial velocities in the northwest and the blue region represents the lower
radial velocities. The third component (grey) seems to correspond to the highest radial velocities but does not show any correlation with the Pearson test.
We also perform a simple linear regression for both components (orange and blue, showed as solid lines in Fig.~\ref{Fig09}$e$), where the correlation is relevant. 

\section{Principal Component Analysis of Data Cubes}\label{sec:PCA}

The basic idea of this analysis, used with multidimensional data, consists in apply a linear orthogonal transformation to take the data from their original basis, where they are correlated, to a new basis, where the variables are not. This new orthogonal
basis, formed by eigenvectors, is then used to represent the data. Eigenvectors in this new coordinate system are classified by decreasing variance \citep{Starck2006}. A tomogram (or eigenimage) is a 2D representation of the projected data in this new basis. Each tomogram corresponds to an eigenspectrum, which is the representation the eigenvector components (or weights) versus the wavelength (or radial velocity). Weight values can be positive or negative which is reflected in the respective tomograms \citep{Cerqueira2015, Steiner2009}.

This variance can be understood as the information quantity contained in each eigenvector. It is possible then to reconstruct the original cube using only the most relevant tomograms (the ones with higher variance), leaving aside 
the rest (basically noise or instrument fingerprint). The use of PCA with data cube has been explored by several authors during the past decade. \citet{Steiner2009} and \citet{Menezes2014} give details about PCA treatment and data preparation, see also \citet{Cerqueira2015} for more details. 
In the following, we are presenting the Principal Component Analysis for both galaxies, first analysing the four more relevant eigenspectra and tomograms and then presenting the reconstructed maps. We choose to perform this decomposition using [O\,{\sc iii}]$\lambda$5007 emission line because it is the brightest line.

\begin{figure*}
\vspace{-0.5cm}
\includegraphics[width=17.0cm, angle=0, scale=1.0]{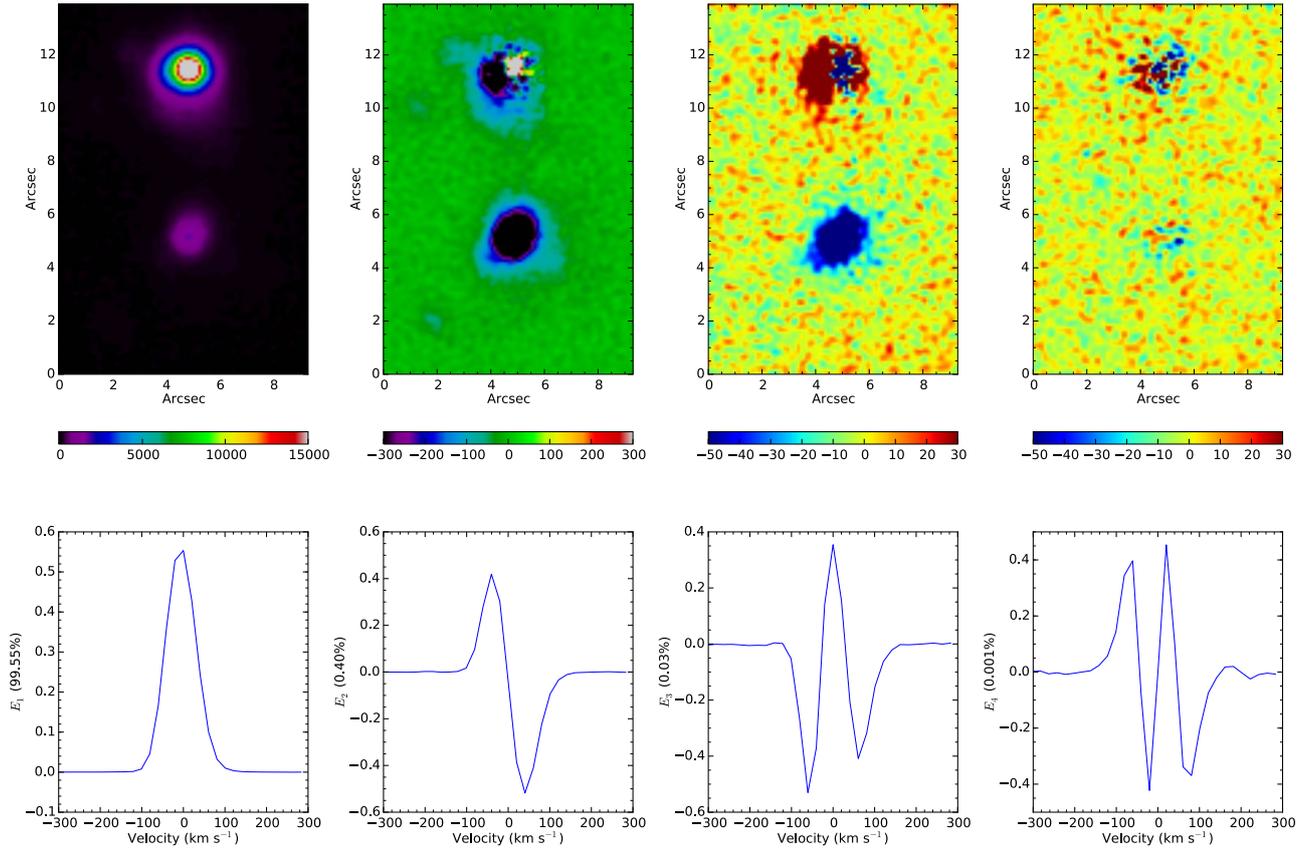}
\caption{UM 461. PCA decomposition of [O\,{\sc iii}]$\lambda$5007 data cube. Top row represents four tomograms with the highest variance. Bottom row presents the correspondent eigenvectors, with the respective variance.}
\label{Fig10}
\end{figure*}

\begin{figure*}
\vspace{-0.5cm}
\includegraphics[width=17.0cm, angle=0, scale=1.0]{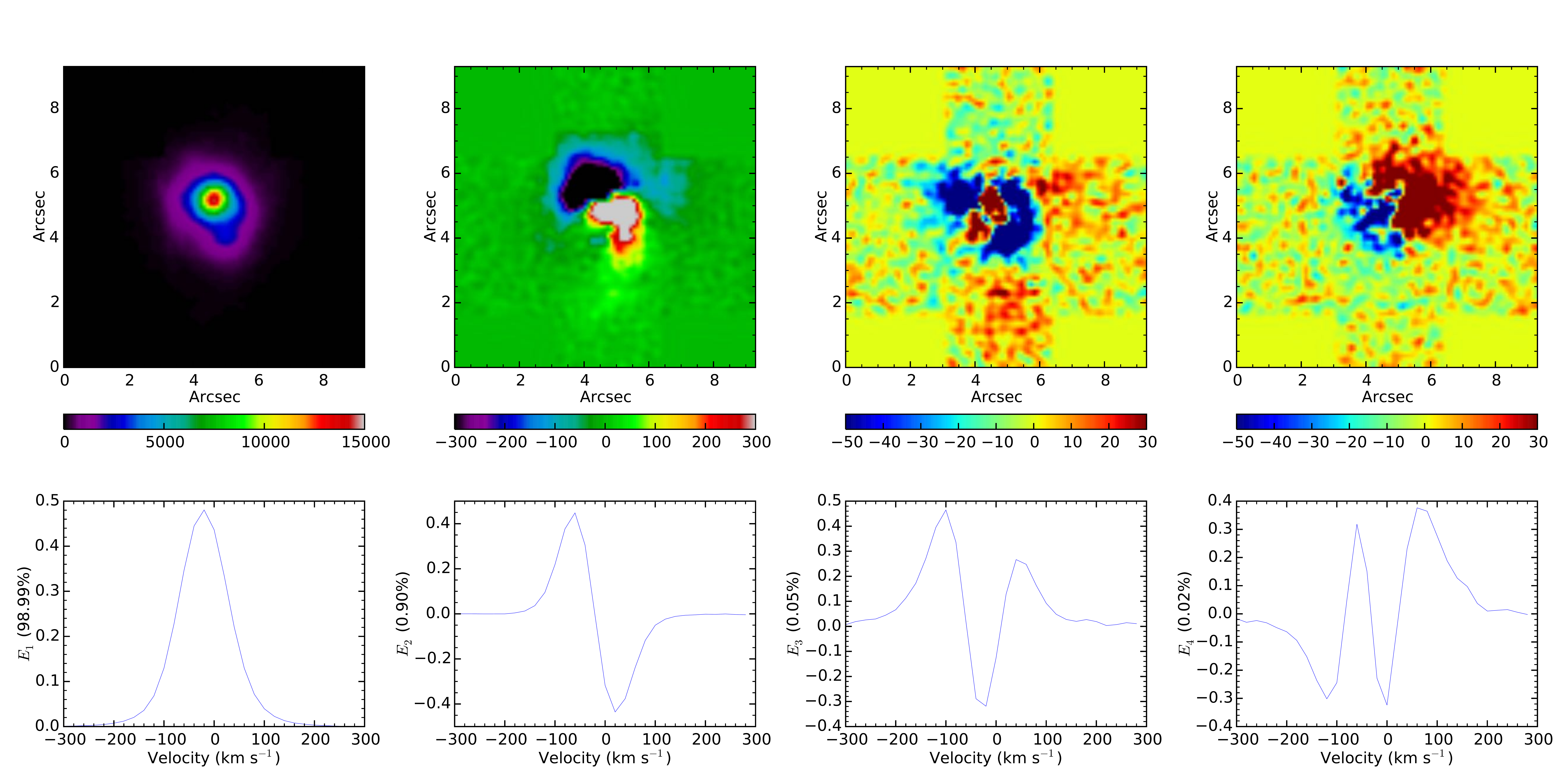}
\caption{CTS 1020. PCA decomposition of [O\,{\sc iii}]$\lambda$5007 data cube. Top row represents four tomograms with the highest variance. Bottom row presents the correspondent eigenvectors, with the respective variance.}
\label{Fig11}
\end{figure*}

\subsection{Tomograms and Eigenspectra}

Fig.~\ref{Fig10} and Fig.~\ref{Fig11}, respectively show the PCA decomposition of UM 461 and CTS 1020. On top, we present the four most relevant tomograms, 
representing respectively 99.55\%, 0.40\%, 0.03\% and 0.001\% of variance for UM 461 and 98.99\%, 0.90\%, 0.05\% and 0.02\% of variance for CTS 1020. For both galaxies, color coding are indicative of the weights (components) of the eigenvectors and are non-dimensional. For both galaxies, eigenspectra are represented in the bottom row. The Y axis represents the weights when the X axis gives the relative displacement, in velocity, with respect to the systemic velocity, respectively v$_{syst}$ = 1039 km~s$^{-1}$ for UM 461 and v$_{syst}$ = 3789 km~s$^{-1}$.

\begin{itemize}

\item The first tomogram, giving most of the variance, mainly represents the monochromatic emission for both galaxies, represented by isophotes in Fig.~\ref{Fig04}$a$ and Fig.\ \ref{Fig05}$a$. We recognise the two nuclei in UM 461 (brighter on the east) and the small extension in the southwest in the case of CTS 1020. \\

\item The second tomogram (and the respective eigenspectrum) shows a different pattern. For both galaxies, we can see that eigenvectors weights are positives and negatives with respect to the velocity displacement.

In the case of UM 461, a positive velocity gradient can be seen with respect to the eigenspectrum, the red wing of the eigenspectrum corresponds to positive weight, associated to positive regions in the East nucleus of the tomogram. The blue wing is associated to the negative region of the East nucleus of the tomogram and with negative weights in the eigenspectrum. The maximum velocity gradient we can infer from the eigenspectrum is $\sim$70 km~s$^{-1}$. The western nucleus doesn't show any gradient.\\

In the case of CTS 1020, positive coefficients appear to be associated with the blueshifted wing of the eigenspectrum and negative coefficients with the redshifted wing. The tomogram associated with the eigenspectrum shows this gradient along a PA $\approx 0^o$. The velocity amplitude of this gradient, deduced from the eigenspectrum, is $\approx80$ km~s$^{-1}$.  \\

\item The third tomogram shows different pattern between UM 461 and CTS 1020. In the case of UM 461, the eigenspectrum shows two negatives peaks (corresponding to blue regions in the tomogram) and one positive peak (corresponding to the red area). CTS 1020 eigenspectrum shows the opposite, two positive peaks and one negative peak. The western nucleus of UM 461 doesn't show velocity gradient and the eastern nucleus is showing  two positive (red) regions with a negative (blue) region in between.  In the case of CTS 1020, two negative (blue) areas with two positive (red) regions in between. \\

\item The fourth tomogram is very faint in UM 461 case but both eigenspectra are showing the same pattern with two negative and two positive peaks.

\end{itemize}

\subsection{Toy Models}

\begin{figure*}
\vspace{-0.5cm}
\includegraphics[width=17.0cm, angle=0, scale=1.0]{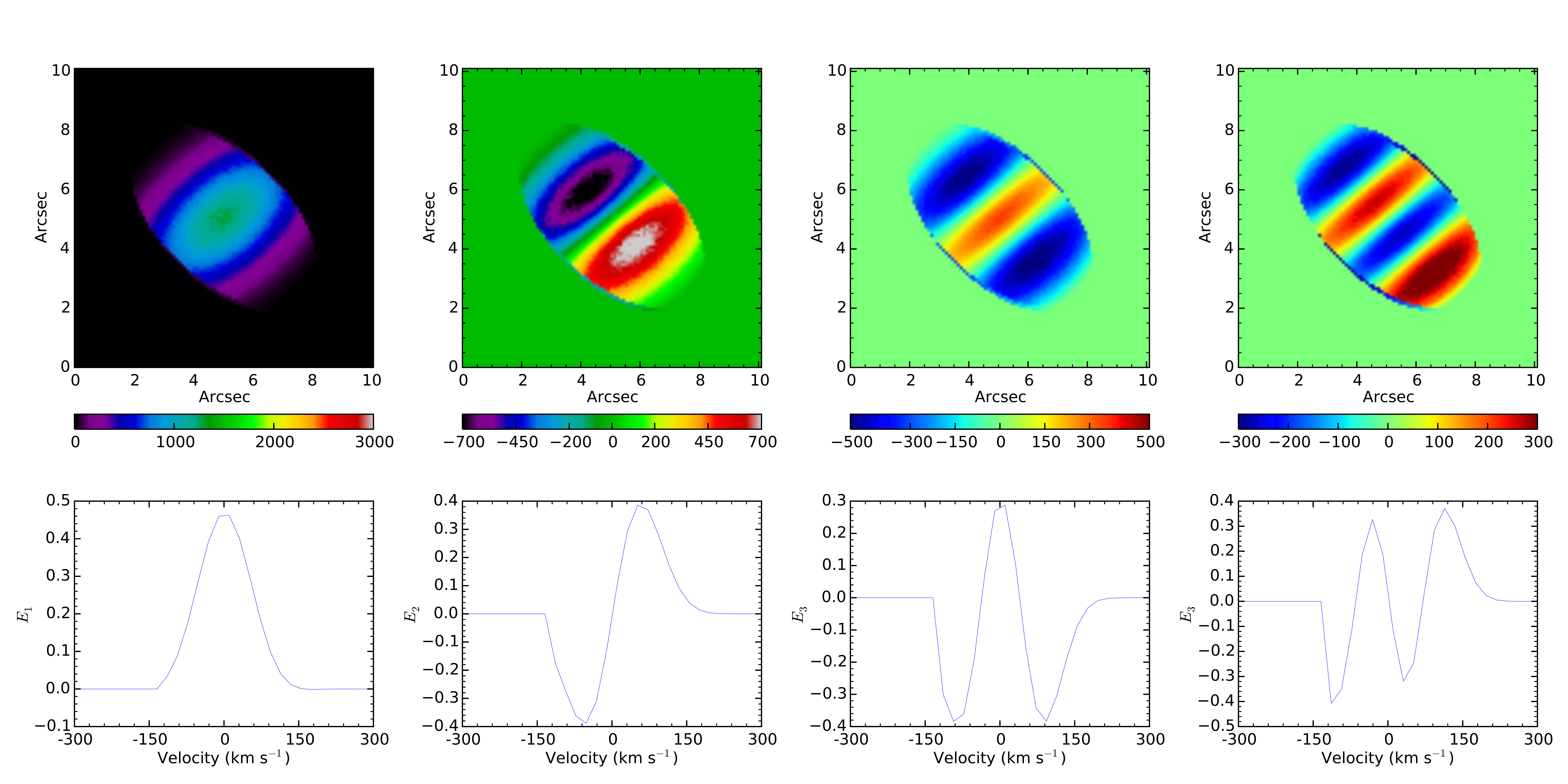}
\caption{PCA decomposition of Rigid Body rotating disk. Top row represents four tomograms with the highest variance. Bottom row presents the correspondent eigenvectors.}
\label{Fig12}
\end{figure*}
\begin{figure*}
\vspace{-0.5cm}
\includegraphics[width=17.0cm, angle=0, scale=1.0]{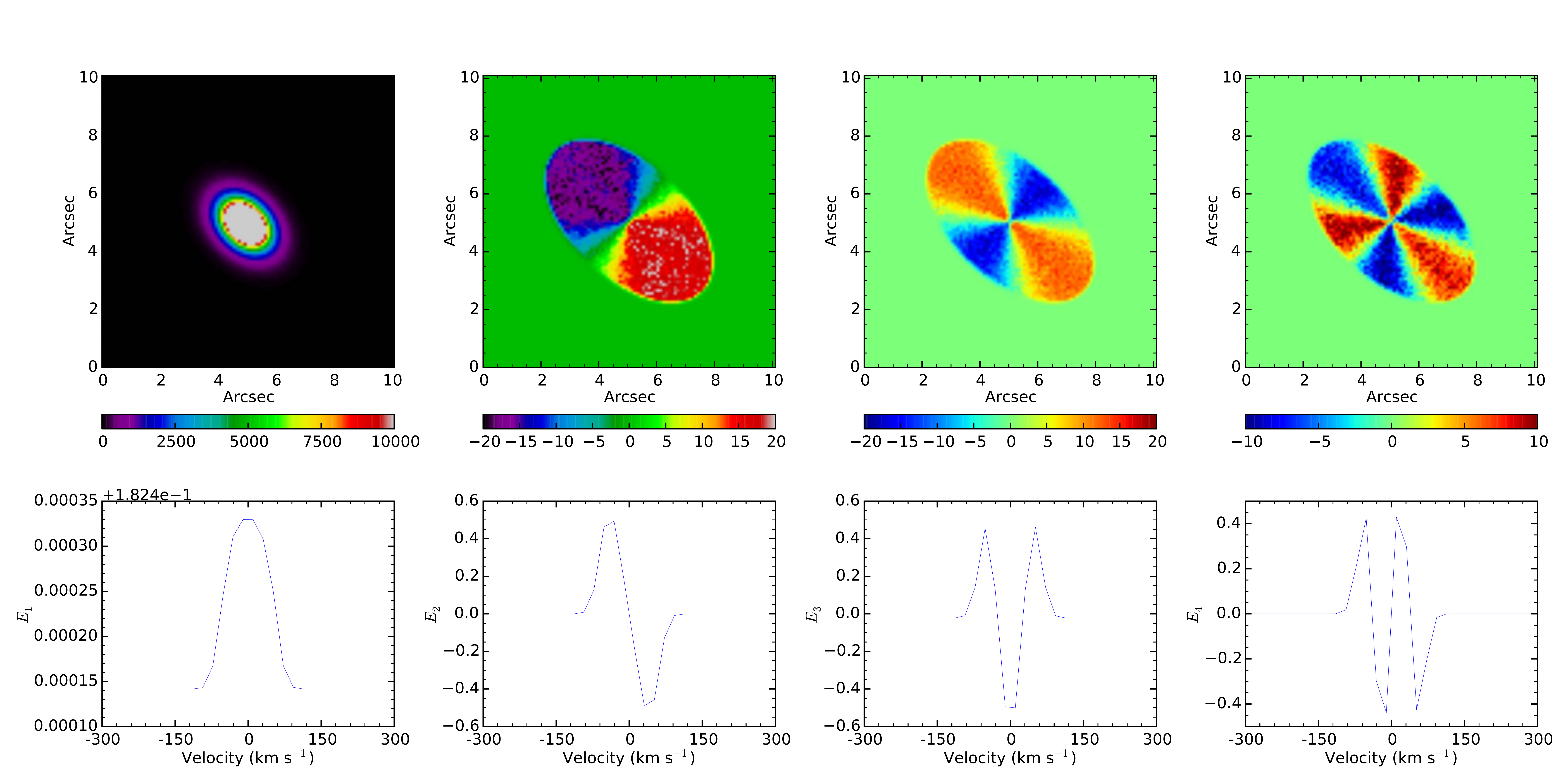}
\caption{PCA decomposition Rotating Disk. Top row represents four tomograms with the highest variance. Bottom row presents the correspondent eigenvectors.}
\label{Fig13}
\end{figure*}

In order to interpret the physical meaning (if any) of eigenspectra and tomograms, we performed some simple toy models to be analysed using PCA and looked at results. Only one phenomenon comes to mind when we think about describing the ionized gas kinematics in a galaxy: a rotating disk. We basically simulate two kinds to rotation
in order to be analysed using PCA, we first present a rigid body rotation and then a differential rotating disk. For the latter one, we decide to give three different inclinations. 
All models have in common an exponential light distribution, a position angle of 45$^o$ and a poissonian noise has been added. The light distribution is not important since it always represents the higher variance in the decomposition and the first tomogram represents it.

\begin{itemize}

\item Rigid Body Rotation \\
Fig.\ref{Fig12} gives the first four tomograms and eigenvectors for the model. As expected, the first component represents the 
exponential light distribution. The second tomogram is the velocity field, with the classic straight line isovelocities. The corresponding eigenvector reproduce the 60 km~s$^{-1}$ velocity gradient. Third and fourth eigenvectors, show alternately positive and negative coefficients. \\

\item Rotating Disk \\ 
In these model we simulate a rotating disk, with three different inclinations, i=20$^0$, i=45$^0$ and i=60$^0$ and a maximum rotational velocity of 60 km~s$^{-1}$. As mentioned before, we found that the second tomogram represents more or less the velocity field (confirmed by the amplitude found in the second tomogram). Fig.\ref{Fig13} only present the result of the PCA for a disk inclination of i=45$^0$. The second tomogram/eigenvector reproduces the velocity field of a rotating disk with a velocity amplitude of 60 km~s$^{-1}$ and a plateau after the first few arc seconds. The third and fourth eigenvectors, also show alternate positive and negative coefficients. The last two tomograms show negative (blue) and positive regions (red).

\end{itemize}  

The principal aim of such simple models was to determine if a PCA decomposition was able to separate some kinematical parameters, such as inclination for instance, and our answer is no. We also simulate rotating disk with two different velocity dispersions, but the different eigenvectors look like the same and no differences can be found.
As a result we are fairly confident that, only the two first eigenvectors/tomograms can have a physical interpretation, respectively as the light distribution (always with the highest variance) and the velocity field of some kind showing a velocity gradient. The interpretation of lower variance of eigenvectors will highly speculative.

\subsection{Reconstructed Data Cubes}

\begin{figure}
\vspace{-0.0cm}
\includegraphics[scale=0.22]{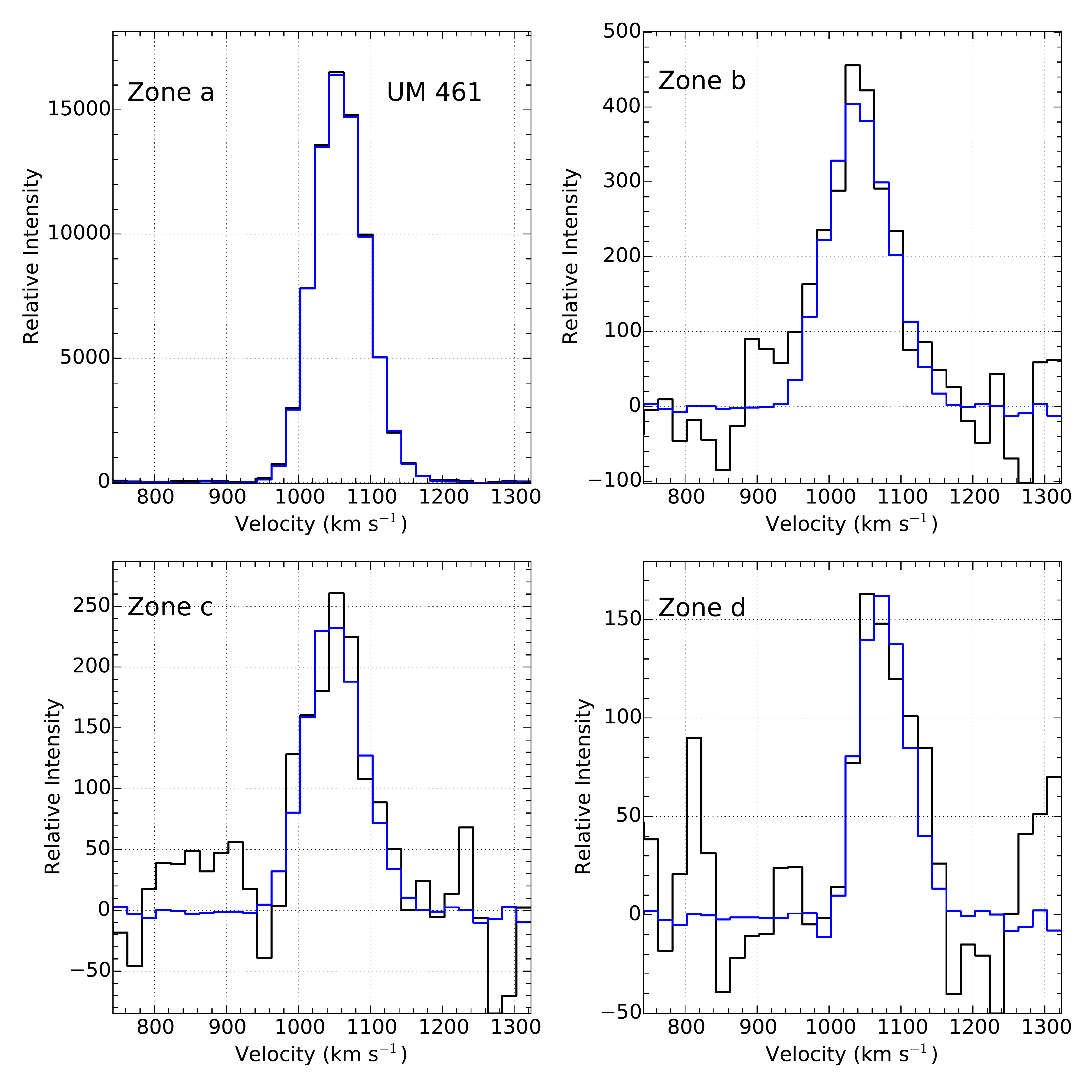}
\caption{UM 461: Reconstructed profiles from the PCA decomposition, in four regions of the galaxy. In blue are represented the reconstructed profile and in black, the original profile.}
\label{Fig14}
\end{figure}
\begin{figure}
\vspace{-0.0cm}
\includegraphics[scale=0.22]{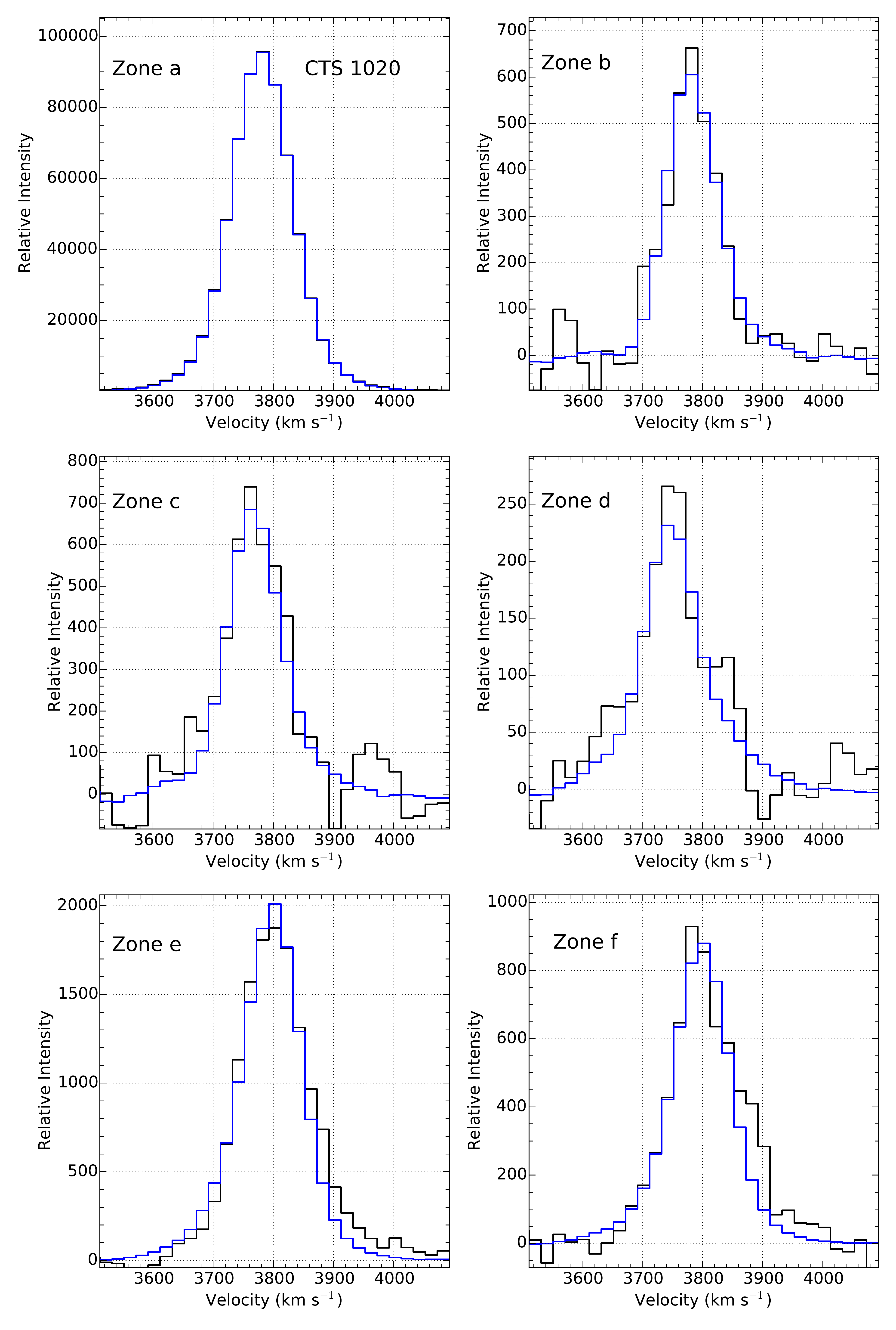}
\caption{CTS 1020: Reconstructed profiles from the PCA decomposition, in four regions of the galaxy. In blue are represented the reconstructed profile and in black, the original profile.}
\label{Fig15}
\end{figure}

The great strength of the PCA decomposition, regarding data reduction, is the fact that it can eliminate noise in a very clean and efficient way. The reconstructed cube will be done by using the most significant eigenvectors (the one with higher variance), following the description by \citet{Steiner2009, Cerqueira2015}. We have limited the reconstruction to the first three eigenvectors representing 99.98\% of the variance in UM 461 case and 99.94\% for CTS 1020. In Fig.~\ref{Fig14} and Fig.~\ref{Fig15}, we present four original and reconstructed profiles in regions across UM 461 and CTS 1020. These regions are $0.4\times 0.4\arcsec$ and are located in different parts of the galaxies. Zones1 are located in the bright parts of each galaxy. The other zones are located in low SNR regions in the outskirts of each galaxy.
In high SNR areas, there are virtually no differences between the original and the reconstructed profile. In the others zones, we can note that reconstructed profiles are more regular in shape. We can also note that the central velocities are almost identical between the original and reconstructed profiles. The major difference comes from the FWHM (and then the velocity dispersion).  In both galaxies, the reconstructed profile appears to be narrower compared to the original profile. We have selected those profiles as examples, where the SNR is low. The differences in velocity dispersion is clearer when looking at the diagnostic diagrams, presented below.

\begin{figure}
\vspace{-0.0cm}
\includegraphics[scale=0.35]{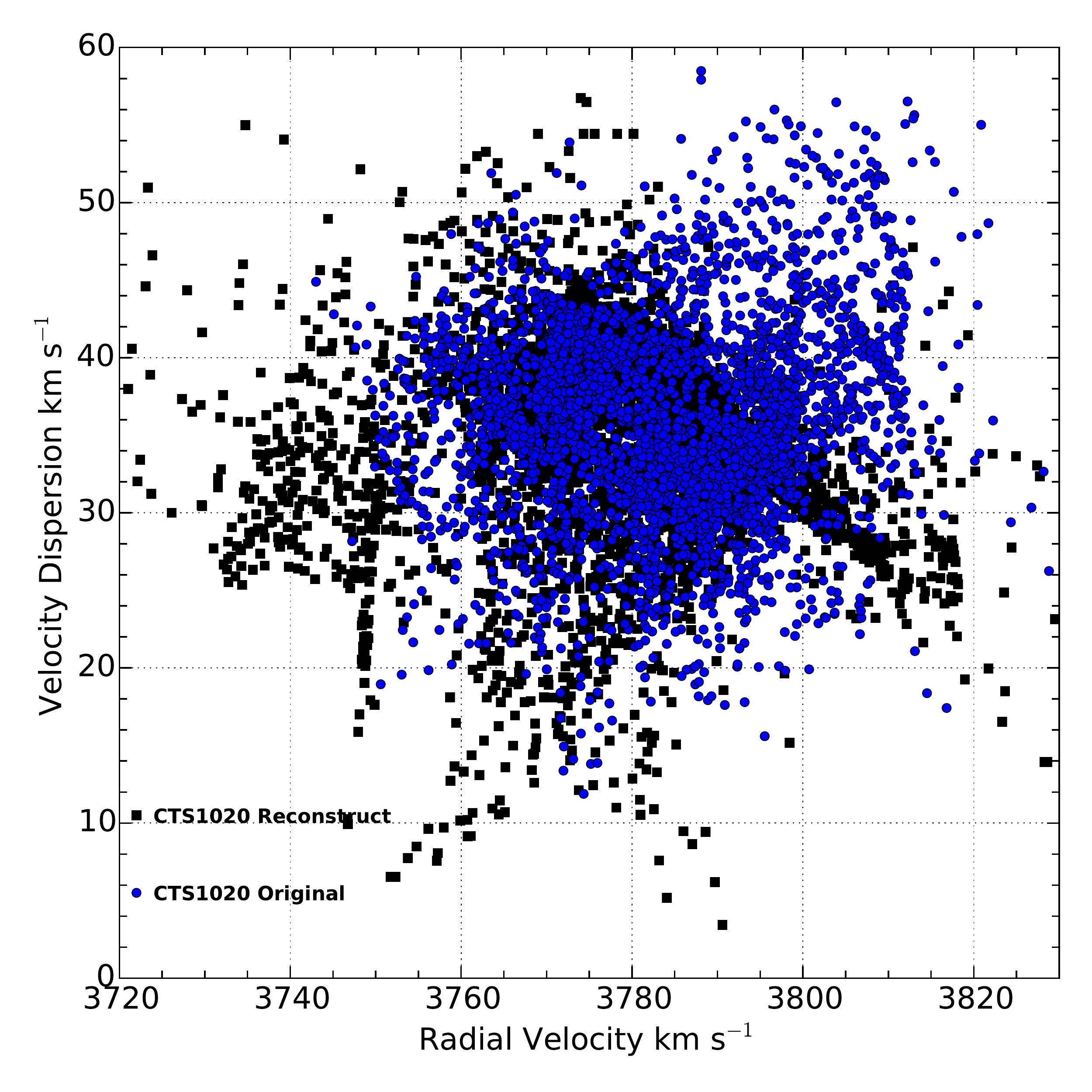}
\caption{Velocity Dispersion vs Radial Velocity diagnostic diagram from reconstructed profiles and original. }
\label{Fig16}
\end{figure}

For both galaxies, we built the different maps and diagnostic diagrams and analysed them. Both the eastern and western parts of UM 461 do not show significant differences in the
$Vr~ vs~ \sigma$, where differences could be seen more easily. Both regions, East and West, show similar distribution in the $Vr~ -~ \sigma$ plane compared to data without PCA analysis. 
In CTS 1020 case, the result is different. Fig.~\ref{Fig16} presents the diagnostic diagram Velocity Dispersion vs Radial Velocity from both reconstructed profiles and originals.  Blue symbols represent original data and black symbols
represent data from reconstructed profiles.
Fig.~\ref{Fig16} shows velocities lower than 3740 km~s$^{-1}$ coming from regions in the South West. Fig.~\ref{Fig16} also shows a lower number of pixels 
with velocity dispersion higher than 45 km~s$^{-1}$ compared to original data. More generally, it is noticeable that the velocity and the velocity dispersion distribution have changed, even if the bulk of points seems to remain the same. The region with high velocity (beyond 3780 km~s$^{-1}$) and high velocity dispersion (larger than 40 km~s$^{-1}$) is not present anymore. These points came from the East - South East region in Fig.~\ref{Fig05}$b$, where the velocity dispersion is the highest.
Profiles from Zones $e$ and $f$ (Fig.~\ref{Fig15}), clearly show that reconstructed profiles are narrower than original ones.
 
\section{Discussion}\label{sec:discussion}

In the following we discuss implications, for both galaxies, of our kinematical study. 

The motion in both star-forming knots of UM 461 has different patterns, suggesting a weak kinematical connection between the knots. The velocity field is disturbed mainly in the western knot, which is spatially resolved in individual stellar clusters and complexes \citep{Noeske2003,Lagos2011}. 
The overall kinematics of the galaxy is probably a result from the interaction with a low mass metal-poor dwarf or H{\,\sc i} cloud occurred \citep{Lagos2018}.

At first glance, CTS 1020 seems to show a more ordered motion with a gradient of $\sim50$ km s$^{-1}$ (from velocity field top to bottom).  The velocity field can be seen as a rotating disc, even though the velocity field is disturbed in several regions.

The velocity dispersion in the eastern knot of UM 461 form a ring-like structure that resembles what we could expect in a collect and collapse scenario \citep{ElmegreenLada1997} in which the first generation of stars produces an H{\,\sc ii} region that expands and sweep up the interstellar medium creating a shell of material that will collapse and start to form stars within it. In that shell, the newly formed stars also create H{\,\sc ii} regions that expands into the shell ionizing it. In that way, the velocity dispersion induced to the gas in the most central region increases as the gas encounter an inhomogeneous medium and increases further as the stars within the shell drive H{\,\sc ii} regions. In that context, the regions of increasing $\sigma$ seen in Fig.\ \ref{Fig04}$c$ could be related to H{\,\sc ii} regions within the ring-like structure. This kind of ring-like structure has been observed in molecular gas that surrounds H{\,\sc ii} regions \citep{Deharveng2003,Deharveng2010}, but in an advanced stage the gas will be ionized by the stellar population formed in the shell and this molecular signature disappears. Thus, we could be looking at this advanced stage of a collect and collapse scenario, also favoured by the morphology of the eastern knot. This is in agreement with \citet{OlmoGarcia2017} that show a global expansion of this knot.

Galaxies also differ in the $I-\sigma$ diagram distribution, but the overall picture still agree with the observed picture for this type of galaxy, in which the $\sigma$ range decreases with increasing intensity. The main difference is that, in UM 461, high intensity regions are related to low velocity dispersion, which is typically observed in H\,{\sc ii} galaxies \citep{Moiseev2012,Moiseev2015}, whereas the high intensity regions in CTS 1020 are related to high velocity dispersion.
Despite this, the main similarity is that gas with low intensity covers the whole $\sigma$ range and permeates the brightest regions of the galaxies.
As proposed by \citet{Moiseev2012}, this pattern is representative of the turbulent motion in the diffuse gas due to the injected mechanical energy from the stellar population. 

On the other hand, in CTS 1020 $\sigma$ decreases along with monochromatic intensity outwards the galaxy, implying that $\sigma$ is driven by virial motion \citep{Moiseev2015}. However, $\sigma$ is also probably affected by 
the injection of mechanical energy from the stellar population that increases the velocity dispersion in the outermost parts of the galaxy, where it should be lower in the proposed scenario. Signatures of disturbed regions can be observed in the velocity dispersion contour map (Fig.\ \ref{Fig05}$c$), where the western and southeastern regions exhibit structures of increasing $\sigma$, and the southeastern where it reaches $\sigma$ higher than 47 km s$^{-1}$.

In summary, by using the $I-\sigma$ diagram along with the velocity dispersion contour map restricted to specific intervals it was possible to separate the line broadening mechanisms. The kinematics in both galaxies is affected by stellar feedback, but in CTS 1020 the gravitational potential dominates, whereas UM 461 seems to be more susceptible to the energy injection from the stellar population. 

The other diagnostic diagram that we used was the $V_r-\sigma$ diagram, which allows to detect systemic motions away or toward us. In order
to find such correlations, we decided to use different statistical tests to separate independent components in this diagram. 

The {\tt MClust} analysis result of the eastern part of UM 461, shows two subcomponents with a weak, but measurable, correlation that points to a motion away from the observer.

In the western knot of UM 461, the {\tt MClust} analysis result also gives two subcomponents, one with weak correlation and the other with no measurable correlation. This analysis points that one region is moving away from us.

The {\tt MClust} analysis of $V_r-\sigma$ diagram for CTS 1020 also reveals three regions, two having weak and moderate correlations.
Those regions appear to reproduce the radial velocity map in Fig.\ \ref{Fig09}$f$. revealing a weak motion toward us.

In summary, the $V_r-\sigma$ diagram analysis for both galaxies shows a correlation between $V_r$ and $\sigma$  compatible with systemic motions toward and/or away from the observer. The use of reconstructed data cubes, after a PCA decomposition, changes the shape of several previously low SNR profile. These changes are reflected in the $V_r-\sigma$ diagram in Fig.\ \ref{Fig16}$c$. 

\citet{Lagos2018} gave a baryonic mass of 1.76~10$^8$ M$_{\odot}$ for UM461. The relation between dynamical mass and baryonic mass for starburst galaxie from \citet{Bergvall2016} gives then an estimated dynamical mass of 3.15~10$^8$ M$_{\odot}$. With UM 461 data: R=0.4kpc and $\sigma$ = 25 km.s$,^{-1}$, we found a dynamical mass (using the same aproximation as \citet{Bergvall2016}) of 2.75~10$^8$ M$_{\odot}$. To reach the predicted dynamical mass, the mean velocity dispersion should be 27 km.s$^{-1}$.

\section{Summary and Conclusions}\label{sec:conclusion}

In this work, we have studied the H{\,\sc ii} galaxies UM 461 and CTS 1020 based on integral field spectroscopy (Gemini GMOS-IFU). Taking advantage of monochromatic, velocity and velocity dispersion maps, we embarque in a kinematical analysis using different diagnostic diagrams (like $I-\sigma$ and $V_r-\sigma$) to investigate the nature of the internal kinematics for both objects. 

As mentioned before, velocity dispersion of ionized gas plays a major role in H{\,\sc ii} galaxies dynamics. The $L~ vs~ \sigma$ relation, based on single measurements, is interpreted as gravity being the main mechanism causing the supersonic broadening of emission profiles \citep{Chavez2014}.

The main result of our study is that the kinematics of ionized gas of these two galaxies is different, but it also shows similarities. Differences come from the velocity and velocity dispersion maps themselves: in UM 461 no ordered motion is present, only velocity gradient; in CTS 1020 a disk like rotation pattern can be seen, even if a larger field of view is necessary to confirm it. 

Velocity dispersion maps show the same differences: in UM 461 regions of low dispersion correspond to high intensity regions, and CTS 1020 shows high dispersion areas where the intensity is the hightest and where intensity is low as well.

$I-\sigma$ diagrams for both galaxies offer some differences. UM 461 diagram shows, according \citet{Moiseev2012}, signature of H\,{\sc ii} regions, constant velocity dispersion and high monochromatic emission, in both knots centers. It also shows the presence of turbulent diffuse gas. 
On the other hand, despite thte fact that, in CTS 1020 case, is still possible to identify the turbulent diffuse gas surrounding the galaxy, high $\sigma$ is related to high intensity and seems to decrease outwards, suggesting that $\sigma$ is driven by virial motions. 

Applying statistical methods, a closer analysis of the $V_r - \sigma$ diagrams, shows that several independent regions with weak and moderate correlation are consistent with systemic motions toward or away the observer. When reported on a geographic map, these regions are consistent with low and high velocities on the velocity maps.  In the case of CTS 1020, it might means that the rotating disk can also be interpreted as large regions animated of opposite motions. A large field of view will be needed in order to find out if the velocity field really represents a rotating disk in that case.

Finally, we also performed a PCA analyis of the data cube in order to improve the SNR. Our results show that data have been improved where the SNR was low, but also show that PCA seems to have modified the shape of the reconstructed profiles, resulting in more symetrical ones.

\section*{Acknowledgments}
This work is Based on observations obtained at the Gemini Observatory, which is operated by the Association of Universities for Research in Astronomy, Inc., under a cooperative agreement with the NSF on behalf of the Gemini partnership: the National Science Foundation (United States), the National Research Council (Canada), CONICYT (Chile), Ministerio de Ciencia, Tecnolog\'{i}$a$ e Innovaci\'{o}n Productiva (Argentina), and Minist\'{e}rio da Ci\^{e}ncia, Tecnologia e Inova\c{c}\~{a}o (Brazil).
MSC would like to thanks Instituto Nacional de Tecnologia - Astrof\'{i}sica for its financial support. HP wants to thanks Casadinho CAPES/Cnpq  project number 552236/2011-0 for its financial support. HP acknowledges the financial support from FAPESB agency under the project number 7916/2015. Authors warmly thanks A.L.B. Ribeiro for his help on the statistical analysis and the use of R. HP thanks the finnancial support from FAPESP agency under the project number 2014/11156-4.
We also acknowledge the usage of the Nasa Extragalactic Database (http://ned.ipac.caltech.edu/) and R free software.





\end{document}